\documentclass{ws-ijmpa}

\begin{document}
  
\markboth{Ji\v{r}\'{i} Kov\'{a}\v{r} and Zden\v{e}k Stuchl\'{i}k}
{Forces in Kerr spacetimes with a repulsive cosmological constant}

%
\catchline{}{}{}{}{}
%

\title{FORCES IN KERR SPACETIMES WITH A REPULSIVE COSMOLOGICAL CONSTANT}

\author{JI\v{R}\'{I} KOV\'{A}\v{R}}

\address{Institute of Physics, Faculty of Philosophy and Science\\
Silesian University, Bezru\v{c}ovo n\'{a}m. 13\\ Opava, 74601,
Czech Republic\\
Jiri.Kovar@fpf.slu.cz}

\author{ZDEN\v{E}K STUCHL\'{I}K}

\address{Institute of Physics, Faculty of Philosophy and Science\\
Silesian University, Bezru\v{c}ovo n\'{a}m. 13\\ Opava, 74601,
Czech Republic\\
Zdenek.Stuchlik@fpf.slu.cz}

\maketitle

\begin{history}
\received{Day Month Year}
\revised{Day Month Year}
\end{history}

\begin{abstract}
Forces defined in the framework of optical reference geometry are introduced in  the case of stationary and axially symmetric Kerr 
black-hole and naked-singularity spacetimes with a repulsive cosmological constant. Properties of the forces acting on test particles moving along circular orbits in the equatorial plane are discussed, whereas it is shown where the gravitational force 
vanishes and changes its orientation and where the centrifugal force vanishes and changes its orientation independently of the 
velocity of test particles related to the optical geometry; the Coriolis force does not vanish for the velocity being non-zero. 
The spacetimes are classified according to the number of circular orbits where the gravitational and centrifugal forces vanish.

\keywords{Kerr-de Sitter spacetimes; gravitational force; centrifugal force.}
\end{abstract}

\ccode{PACS numbers: 04.70.-s, 04.25.-g, 04.20.-q}

\section{Introduction}

In general relativity, the description of particle motion is usually not so illustrative as in the classical physics, where the Newtonian forces can be routinely used. However, using the optical reference geometry given by M. Abramowicz and his co-workers 
\cite{Abr-Car-Las:1988:GENRG2:,Abr:1990:MONNR:,Abr-Nur-Wex:1993:CLAQG:,Abr-Nur-Wex:1995:CLAQG:},       we can define forces which surprisingly enables us to describe the relativistic dynamics in accordance with our natural Newtonian intuition. Moreover, many important properties of relativistic dynamics in terms of the forces can be effectively illustrated by the 
properties of embedding diagrams of the optical reference geometry \cite{Kri-Son-Abr:1998:GENRG2:,Stu-Hle:1999:ACTPS2:,Stu-Hle:1999:CLAQG:,Stu-Hle-Jur:2000:CLAQG:,Hle:2001:RAGtime2and3:}. These are just the main advantages of the forces defined by M. Abramowicz compared to a number of other forces defined in the framework of general relativity \cite{Jan-Car-Bin:1992:ANNPH1:}. 
 
The forces related to the optical reference geometry were exploited in many papers in the case of various types of spacetimes \cite{Abr-Mil:1990:MONNR:,Stu:1990:BULAI:,Abr:1992:MONNR:,Abr-Mil-Stu:1993:PHYSR4:,Abr:1993:RenGRCos:,Abr-etal:1997:CLAQG:,Abr-etal:1997:GENRG2:,Abr:1999:PHYSR:}.
Here we consider physical relevant stationary and axially symmetric backgrounds around rotating black holes or naked singularities in the universe with recently indicated repulsive cosmological constant, i.e., Kerr-de~Sitter (KdS hereinafter) spacetimes. Black-hole spacetimes with a non-zero cosmological constant were treated in terms of the optical reference geometry in the simplest, 
spherically symmetric cases of the Schwarzschild-de~Sitter (SdS hereinafter) and Reissner-Nordstr\"{o}m-de Sitter spacetimes
\cite{Stu:1990:BULAI:,Stu-Hle:1999:PHYSR4:,Stu-Hle:2002:ACTPS2:}. In the later case, naked-singularity spacetimes appear along with 
black-hole spacetimes. 

In Section 2, a brief summary of the Abramowicz definition of the forces in the framework of optical reference geometry is given. It 
is shown that in the special reference frame of general stationary spacetimes, the motion of test particles can be described by 
using gravitational, centrifugal, Coriolis, and Euler forces. The behavior of these forces can be elucidated in many cases 
of the test particle motion. In this paper, we focus on the forces acting on particles in stationary motion along circular orbits in 
the equatorial plane of the stationary, axially symmetric KdS spacetimes. In Section 3, we give general form of the 
forces (naturally, the Euler force vanishes in the case of the stationary circular motion) and in Section 4, we summarize basic 
features of the KdS geometry. The fundamental properties of the forces are discussed in Section 5. We show that in a given 
spacetime, there are several circular orbits where the gravitational force vanishes and changes its orientation and, generally, 
several other orbits where the centrifugal force vanishes and changes its orientation independently of the velocity of test particles. 
On the other hand, the Coriolis force does not vanish for non-zero velocity. We also discuss the vanishing and change of orientation 
of the forces in the limit cases of the KdS spacetimes, i.e., in the Kerr and SdS spacetimes, which
clearly demonstrates the influence of the cosmological constant and rotation of the central object on moving test particles. In Section 6, some concluding remarks are presented.

\section{Forces in stationary spacetimes}

In a stationary spacetime described by a metric $g_{ik}$ (with signature +2), we can assume a family of special observers with a timelike and unit 4-velocity field $n^{i}$ and with its \mbox{4-acceleration} field equal to the gradient of a scalar function called 'gravitational potential', i.e., \cite{Abr-Nur-Wex:1993:CLAQG:}
\begin{eqnarray}
\label{1}
n^k n_k=-1,\quad
n^i\nabla_i n_k=\nabla_k\Phi.
\end{eqnarray}
It is useful (but not necessary for the definition of the 
forces) to require the vector field $n^i$ to satisfy the condition of hypersurface orthogonality
\begin{eqnarray}
\label{3}
n_{[i}\nabla_j n_{k]}=0.
\end{eqnarray}
We can find at least two solutions of the equations~(\ref{1}); $\Phi=const$ and $n^i$ corresponding to the \mbox{4-velocity} field of freely 
falling (geodesic) observers or $\Phi=\frac{1}{2}\ln{(-\iota^i\iota_i)}$ and $n^i=e^{-\Phi}\iota^i$ corresponding 
to the \mbox{4-velocity} field of stationary observers. In the second case, $n^i$ is the unit vector field parallel 
to the timelike Killing vector field $\iota^i$, which exists due to the spacetime stationarity.
The equation  
\begin{eqnarray}
\label{4}
n^i\nabla_{i}\Phi=0,
\end{eqnarray}
following from the equations~(\ref{1}), suggests that the special observers with the \mbox{4-velocity} $n^i$ observe no change in the 
gravitational potential as their proper time passes. It means the observers are fixed with respect to the \textit{gravitational field}.
The local instantaneous \mbox{3-dimensional} ($n^i$ orthogonal) space of the observers is described by the metric
\begin{eqnarray}
\label{5}
h_{ik}=g_{ik}+n_i n_k,
\end{eqnarray}
the so-called \textit{directly projected geometry}. It is useful to define the conformally adjusted metric of the spacetime
\begin{eqnarray}
\label{6}
\tilde{g}_{ik}=e^{-2\Phi}g_{ik}
\end{eqnarray}
and the conformally adjusted metric of the directly projected geometry
\begin{eqnarray}
\label{7}
\tilde{h}_{ik}&=&e^{-2\Phi}h_{ik},
\end{eqnarray}
the so-called \textit{optical reference geometry}.   

The \mbox{4-velocity} $u^i$ of a test particle with a rest mass $m$ can be decomposed into the time part and the spatial part 
($n^i$ orthogonal) in the reference frame of the special observers with the \mbox{4-velocity} $n^i$ by using the relation
\begin{eqnarray}
\label{8}
u^i=\gamma(n^i+v\tau^i),
\end{eqnarray}
where $\tau^i$ is the unit spacelike vector ($n^i$ orthogonal) along with the \mbox{3-velocity} with the magnitude $v$ (measured by special observers) is aligned. The quantity $\gamma=(1-v^2)^{-1/2}$ is the Lorentz factor, i.e., the normalization factor that makes $u^iu_i=-1$. Moreover, there are $\gamma v\tau^i=u^k h^i_k$, $\gamma=-n^i u_i$, and $h^i_k=\delta^i_k+n^i n_k$ is the projection tensor allowing 
the special observer to define \mbox{3-dimensional} quantities by projecting \mbox{4-dimensional} quantities into his local 
instantaneous \mbox{3-dimensional} space.

The \mbox{4-acceleration} of the particle is defined by the relation $a_k=u^i\nabla_i u_k$. Using the relation~(\ref{8}), we can 
easily derive the following formula for the \mbox{4-acceleration} (in which terms are arranged according to the powers of the 
velocity and its derivate)
\begin{eqnarray}
\label{9}
a_k=\gamma^2\nabla_k\Phi +\gamma^2 v(n^i\nabla_i\tau_k+\tau^i\nabla_{i}n_k)+\gamma^2v^2\tau^i\nabla_i\tau_k+
(v\gamma\dot{)}\tau_k+\dot{\gamma}n_k,
\end{eqnarray}
where $(v\gamma\dot{)}=u^i\nabla_i(\gamma v)$.
By using the spacelike unit vector in the optical reference geometry parallel to $\tau^i$, i.e., the vector 
$\tilde{\tau}^{i}=e^{\Phi}\tau^{i}$, its covariant form $\tilde{\tau}_i=e^{-\Phi}\tau_i$, the scalar $E=-\iota^iu_i$, the 
identity $\gamma^2=1+v^2\gamma^2$ and by denoting $\tilde{v}=\gamma v$, we obtain (after amount of simple but tedious algebra) 
the \mbox{4-acceleration} in the form 
\begin{eqnarray}
\label{10}
a_k=\nabla_k\Phi +\gamma^2v(n^i\nabla_i\tau_k+\tau^i\nabla_in_k)+\tilde{v}^{2}\tilde{\tau}^i\tilde{\nabla}_i\tilde{\tau}_k+
(Ev\dot{)}\tilde{\tau}_k+\dot{\gamma}n_k,
\end{eqnarray}
where $\tilde{\nabla}_i$ denotes the covariant derivate with respect to the optical reference geometry. Note that $\tau^i$ is defined only along the world line of the particle. But in our construction, it is necessary to know how $\tau^i$ changes along $n^i$. This can be done in several different ways (different gauges). The gauge used by M. Abramowicz and his co-workers assures that the Lie derivate of $\tau^i$ with respect to $\iota^i=e^\Phi n^i$ vanishes, i.e., $\mathcal{L}_\iota\tau_k\equiv \iota^i\nabla_i \tau_k-\tau^i\nabla_i\iota_k=0$, which causes that the Coriolis  force defined below vanishes in static spacetimes \cite{Abr-Nur-Wex:1993:CLAQG:}. 

By projecting the \mbox{4-acceleration}~(\ref{10}) into the \mbox{3-dimensional} space and by using the 
condition of hypersurface orthogonality~(\ref{3}), we arrive at the formula 
\begin{eqnarray}
\label{11}
a_j^{\perp}&=&h^k_ja_k=\nabla_j\Phi+\tilde{v}^{2}\tilde{\tau}^i\tilde{\nabla}_{i}\tilde{\tau}_{j}+\gamma^2vX_{j}+\dot{V}\tilde{\tau}_j, 
\end{eqnarray}
where $X_j=n^i(\nabla_i\tau_j-\nabla_j\tau_{i})$ and $\dot{V}=u^i\nabla_i(Ev)$.

A real force acting on a particle (for example a thrust of rocket orbiting a black hole) in the \mbox{3-dimensional} space can 
be expressed by the relation $F_k^{\perp}=ma_k^{\perp}$. This equation can be rewritten to the form $F_k^{\perp}-ma_k^{\perp}=0$, 
which suggests that the particle is not accelerated in its own comoving frame and the real force is balanced by the inertial force 
${F_k'}^{\perp}=-m a_k^{\perp}$,  i.e., ${F_k}^{\perp}+{F'_k}^{\perp}=0$. Due to the formula~(\ref{11}), we can decompose this 
inertial force into the sum of the gravitational ${G_k}^{\perp}$, centrifugal ${Z_k}^{\perp}$, Coriolis 
${C_k}^{\perp}$, and Euler ${E_k}^{\perp}$ forces, which are familiar from the Newtonian physics
\begin{eqnarray}
\label{12}
{F'_k}^{\perp}=-ma_k
^{\perp}={G_k}^{\perp}+{Z_k}^{\perp}+{C_k}^{\perp}+{E_k}^{\perp},
\end{eqnarray}
where
\begin{eqnarray}
\label{13}
G_k^{\perp}&=&-m\nabla_k\Phi,\\
\label{14}
Z_k^{\perp}&=&-m\tilde{v}^2\tilde{\tau}^i\tilde{\nabla}_{i}\tilde{\tau}_k,\\
\label{15}
C_k^{\perp}&=&-m\gamma^2vX_k,\\
\label{16}
E_k^{\perp}&=&-m\dot{V}\tilde{\tau}_k.
\end{eqnarray}
Note that in the context of the optical geometry relativistic 
approach, the gravitational force ranks among the inertial forces. 

\section{Circular motion in stationary and axially symmetric spacetimes}

Stationary and axially symmetric spacetimes admit two Killing vector fields: the timelike vector field $\mathbf{\eta}=\partial/\partial{t}$ and the spacelike vector field $\mathbf{\xi}=\partial/\partial{\phi}$. These Killing vector fields are not orthogonal in general and $\eta^i\eta_{i}=g_{tt}$, $\eta^i\xi_i=g_{t\phi}$, $\xi^i\xi_i=g_{\phi\phi}$. 
Their combination, especially $\iota^i=\eta^i+\Omega_{LNRF}\xi^i$, where $\Omega_{LNRF}=-\eta^i\xi_i/\xi^i\xi_i$, can be used for the definition of the special observers 
\begin{eqnarray}      
\label{30}
n^i&=&e^{-\Phi}(\eta^i+\Omega_{LNRF}\xi^i),\\
\Phi&=&\frac{1}{2}\ln{[-(\eta^i+\Omega_{LNRF}
\xi^i})(\eta_{i}+\Omega_{LNRF}
\xi_i)].
\end{eqnarray}
This \mbox{4-velocity}, relevant for the construction of the ordinary projected geometry, 
corresponds to the \mbox{4-velocity} of the locally non-rotating frames moving along circular orbits with the angular velocity $d\phi/dt=\Omega_{LNRF}$. The timelike vector field (\ref{30}) is the unit and hypersurface orthogonal vector field, whereas its \mbox{4-acceleration} equals to the gradient of the scalar function $\Phi$, just as required in the equations~(\ref{1}) and~(\ref{3}).    
On the other hand, the \mbox{4-velocity}     
\begin{eqnarray}
\label{31}
u^i&=&e^{-A}(\eta^i+\Omega\xi^i),\\
A&=&\frac{1}{2}\ln{[-(\eta^i+\Omega
\xi^i})(\eta_{i}+\Omega
\xi_i)]
\end{eqnarray}
corresponding to the circular motion of test particle with the angular velocity $d\phi/dt=\Omega\neq\Omega_{LNRF}$ is not hypersurface orthogonal, i.e., it does not satisfy the condition~(\ref{3}). Just this property of the \mbox{4-velocity} of special observers, i.e., $n^i$ being  hypersurface orthogonal, naturally ensures that the unit vector $\tau^i$, used in the decomposition~(\ref{8}), is located in the hypersurface. Moreover, due to the circular orbits, it is directed along the Killing vector $\xi^i$, i.e.,
\begin{eqnarray}
\label{32}
\tau^i=(\xi^k\xi_k)^{-1/2}\xi^i.
\end{eqnarray}
Thus by using relations~(\ref{31}) and~(\ref{8}), we obtain the Lorentz factor and the velocity in the form
\begin{eqnarray}
\label{33}
\gamma&=&e^{\Phi-A},\\
\label{34}
v&=&e^{-\Phi}(\xi^{i}\xi_{i})^{1/2}(\Omega-\Omega_{LNRF}).
\end{eqnarray}
Note that $\Omega=\Omega_{LNRF}$ is equivalent to $\gamma=1$ and $v=0$. 

Using the general forms of the inertial forces~(\ref{13})-(\ref{16}) (dropping the superscript~$^{\perp}$)  and the 
relations~(\ref{31})-(\ref{34}), we arrive at the expressions for components of the gravitational, centrifugal, and
Coriolis forces acting on the particle moving along the circular orbit with $\Omega=\mathrm{const}$
\begin{eqnarray}
\label{35}
G_k&=&-m\,\frac{1}{2}\nabla_{k}(\ln{e}^{2\Phi}),\\
\label{36} 
Z_k&=&m(\gamma{v})^{2}\,\frac{1}{2}(\xi^i\xi_i)^{-1}e^{-2\Phi}[e^{2\Phi}\nabla_{k}{(\xi^i\xi_i)}-{\xi^i\xi_i}{\nabla_k e^{2\Phi}}],\\
\label{37}
C_k&=&m\gamma^{2}v\, (\xi^i\xi_i)^{-3/2}e^{-\Phi}[\xi^i\xi_i\nabla_{k}{(\eta^i\xi_i)}-\eta^i\xi_i\nabla_k(\xi^i\xi_i)].
\end{eqnarray}
Note that the Euler force $E_k$ is non-zero in the case of $\Omega\neq\mathrm{const}$, being determined by $\dot{\Omega}=u^i\nabla_i\Omega$. Due to the axial symmetry and stationarity of the spacetimes, the $t$ and $\phi$-components vanish. We will focus on the inertial forces acting on particles moving only in the equatorial plane~($\theta=\pi/2$) where also the $\theta$-components vanish. The gravitational force is velocity-independent, the centrifugal force depends on the second power of the velocity, and the Coriolis force depends on the first power of the velocity. This indicates the Newtonian character of the forces. Moreover, the acceleration necessary to keep a particle in a stationary motion with a velocity \mbox{$v$} along a circular orbit $r=const$ in the equatorial plane can be expressed in a very simple form 
\begin{eqnarray}
\label{41a}
a&=&-\mathcal{G}-(\gamma v)^2\mathcal{Z}-\gamma^2 v\mathcal{C},
\end{eqnarray}
where 
\begin{eqnarray}
\label{42}
\mathcal{G}&=&\frac{G_r}{m}=-\frac{1}{2}\nabla_{r}(\ln{e}^{2\Phi})=-\nabla_{r}\Phi,\\
\label{43}
\mathcal{Z}&=&\frac{Z_r}{m(\gamma{v})^{2}}=\frac{1}{2}(\xi^i\xi_i)^{-1}e^{-2\Phi}[e^{2\Phi}\nabla_{r}{(\xi^i\xi_i)}-\xi^i\xi_i\nabla_{r}e^{2\Phi}],\\
\label{44}
\mathcal{C}&=&\frac{C_r}{m\gamma^{2}v}=(\xi^i\xi_i)^{-3/2}e^{-\Phi}[\xi^i\xi_i\nabla_{r}(\eta^i\xi_i)-\eta^i\xi_i\nabla_{r}(\xi^i\xi_i)]
\end{eqnarray}
are mass and velocity independent parts of the $r$-components of the forces. Thus, analogous to the Newtonian dynamics, this equation enables an effective discussion of the properties of both accelerated and geodesic motion~\cite{Stu-Hle-Jur:2000:CLAQG:}. 

\section{Kerr-de~Sitter spacetimes}

KdS spacetimes are stationary and axially symmetric solutions of Einstein's vacuum equations with a positive (repulsive) 
cosmological constant. The KdS solution describes geometry of spacetime around an isolated Kerr (rotating and uncharged) 
black hole or naked singularity determined by its mass $M$ and specific angular momentum $a$ in the universe with the repulsive 
cosmological constant $\Lambda>0$. 

In the standard Boyer-Lindquist coordinates $(t,r,\theta,\phi)$ and geometric units $(c=G=1)$, the line element of the KdS 
geometry is given by the relation
\begin{eqnarray}
\label{45}
ds^{2}=-\frac{\Delta_r}{I^2\rho^2}(dt-a\sin^2{\theta}d\phi)^2+\frac{\Delta_{\theta}\sin^2{\theta}}{I^2\rho^2}[adt-(r^2+a^2)d\phi]^2+\\\nonumber
\frac{\rho^2}{\Delta_r}dr^2+\frac{\rho^2}{\Delta_{\theta}}d\theta^2,
\end{eqnarray}
where
\begin{eqnarray}
\label{46}
\Delta_r&=&r^2-2Mr+a^2
    -\frac{1}{3}{\Lambda}r^2(r^2+a^2),\\
\label{47}
\Delta_{\theta}
  &=&1+\frac{1}{3}{\Lambda}a^2\cos^2{\theta},\\
\label{48}
I&=&1+\frac{1}{3}{\Lambda}a^2,\\
\label{49}
\rho^2&=&r^{2}+a^2\cos^{2}{\theta}. 
\end{eqnarray}
It is convenient to use the following dimensionless quantities: $r/M\rightarrow r$, $t/M\rightarrow t$, $s/M\rightarrow s$, 
$a/M\rightarrow a$ and introduce the dimensionless cosmological parameter
\begin{eqnarray}
\label{50}
y=\frac{1}{3}\Lambda M^2,
\end{eqnarray}
i.e., we express all of these quantities in units of $M$. It is equivalent to putting $M=1$. 

The KdS geometry, being stationary and axially symmetric, admits the Killing vector fields  
$\mathbf{\eta}=\partial/\partial{t}$ and $\mathbf{\xi}=\partial/\partial{\phi}$. 
The stationary regions of the spacetimes are determined by the relation $\Delta_r(r;a^2,y)\geq 0$ and these are 
limited by the inner and outer black-hole horizons at $r_{h-}$ and $r_{h+}$ and by the cosmological horizon at $r_{c}$, which are the real roots of the equation $\Delta_r(r;a^2,y)=0$. Then spacetimes containing three horizons are black-hole spacetimes, 
while spacetimes containing one horizon (the cosmological horizon exists for any choice of the spacetime parameters) are 
naked-singularity spacetimes. Spacetimes with two horizons are extreme black-hole or extreme  naked-singularity spacetimes. 

\begin{figure}[pb]
\centerline{\psfig{file=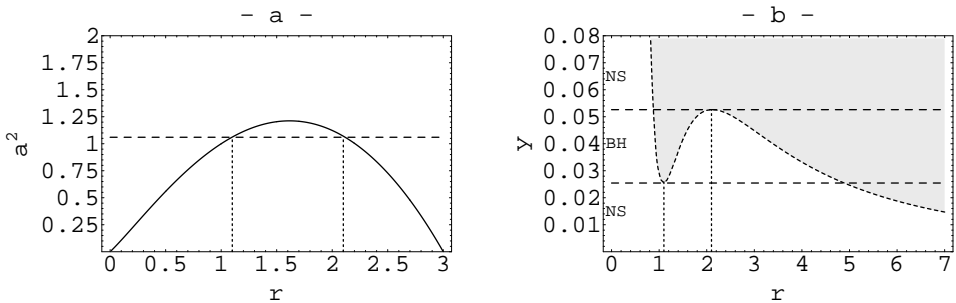,width=1.1\hsize}}
\vspace*{8pt}
\caption{(a) The characteristic function $a^2_{he}(r)$ governing the location of extrema of the function $y_h(r;a^2)$. For given $a^2$, the extrema of $y_h(r;a^2)$ are determined by the solutions of $a^2=a^2_{he}(r)$ (note the dashed line $a^2=1.06$ and compare with Fig. \ref{Fig:1}b).     
(b) The function $y_h(r;a^2)$ determining the loci of event horizons of the KdS spacetimes and limiting the dynamic region where $y>y_h(r;a^2)$ (gray). The function is given for $a^2=1.06$. For given $y$ and $a^2$, the horizons are determined by the solutions of $y=y_h(r;a^2)$. In the parameter line $(y)$ of the spacetimes, the extrema of $y_h(r;a^2)$  separate regions corresponding to the black-hole spacetimes BH and naked-singularity spacetimes NS for given values of the rotational \mbox{parameter $a^2$}.}   
\label{Fig:1}
\end{figure}

It follows from the relation $\Delta_r(r;a^2,y)=0$ that for given values of the rotational and cosmological parameters $a^2$ and $y$, 
the loci of horizons are given by solutions of the equation  
\begin{eqnarray}
\label{54}
y=y_h(r;a^2)
\equiv\frac{r^2-2r+a^2}{r^2(r^2+a^2)}.
\end{eqnarray}
Because of the repulsive cosmological constant, the solutions are restricted by the condition 
\begin{eqnarray}
\label{55}
y_h(r;a^2)>0.
\end{eqnarray} 
The asymptotic behavior of the function $y_h(r;a^2)$ is given by $y_h(r\rightarrow\infty;a^2)\rightarrow+0$ and $y_h(r\rightarrow 0;a^2)\rightarrow \infty$. The local extrema of  $y_h(r;a^2)$ are determined (due to the condition $\partial_r\: y_h(r;a^2)=0$) by the relation
\begin{eqnarray}
\label{58}
a^2(r)=a^2_{he}(r)\equiv\frac{1}{2}(-2r^2+\sqrt{8r+1}r+r).
\end{eqnarray}
The maximum of the function $a^2_{he}(r)$ is located at $r\doteq1.616$ and takes the value $a^2_{he,max}\doteq1.212$ (see Fig.~\ref{Fig:1}a).

We can distinguish three different types of behavior of $y_h(r;a^2)$ (see Fig.~\ref{Fig:3}).
For $a^2<1.212$, $y_h(r;a^2)$ has two local extrema $y_{h,min}(a^2)$ and $y_{h,max}(a^2)$ determined by~(\ref{54}) and~(\ref{58}) 
(see Fig.~\ref{Fig:1}b). The black-hole spacetimes exist for $y_{h,min}(a^2)\leq y< y_{h,max}(a^2)$ and $y>0$, while the naked-singularity 
spacetimes exist for $0<y<y_{h,min}(a^2)$ or $y\geq y_{h,max}(a^2)$. For $a^2=1.212$, the extrema $y_{h,min}(a^2)$ and $y_{h,max}(a^2)$ 
coincide at $y_{h,crit}=0.059$, which is the limiting value for the black-hole spacetimes. For $a^2>1.212$, $y_h(r;a^2)$ has no 
extrema and there are only the naked-singularity spacetimes \cite{Stu-Sla:2004:PHYSR4:}. The parameter plane $(a^2,y)$ 
separated by the functions $y_{h,min}(a^2)$ and $y_{h,max}(a^2)$ into the regions corresponding to the black-hole and 
naked-singularity spacetimes is illustrated in Figs~\ref{Fig:4}-\ref{Fig:5}.   

\section{Forces in Kerr-de~Sitter spacetimes}

We will discuss the vanishing and change of orientation of the gravitational, centrifugal, and Coriolis forces acting on a particle in 
stationary motion along a circular orbit in the equatorial plane of the KdS spacetimes.  
The velocity dependent forces (centrifugal and Coriolis) naturally vanish for $v=0$. It is, however much more interesting to establish 
the radii of circular orbits where the forces vanish independently of the velocity of the motion. It means, we have to determine the zero points of the velocity independent parts of the forces~(\ref{42})-(\ref{44}), which in KdS spacetimes with the line element~(\ref{45}) take the form
\begin{eqnarray}
\label{61}
\mathcal{G}(r;a^2,y)=\{r\Delta_r[ra^4y+(yr^3+r+2)a^2+r^3]\}^{-1}\{r^3a^2(a^2+r^2)^2y^2+\nonumber\\
r^2(a^2+r^2)[r^3+a^2(r+4)]y-r^4-2r^2a^2+4ra^2-a^4\},
\end{eqnarray}
\begin{eqnarray}
\label{62}
\mathcal{Z}(r;a^2,y)=\{r\Delta_r[ra^4y+(yr^3+r+2)a^2+r^3]\}^{-1}\{r^3a^4(a^2+r^2)y^2+\nonumber\\
r^2a^2[(2r+5)a^2+r^2(2r+3)]y+r^4(r-3)+\nonumber\\
ra^2[r(r-3)+6]-2a^4\},
\end{eqnarray}
\begin{eqnarray}
\label{63}
\mathcal{C}(r;a^2,y)=\frac{2a(a^2+3r^2)}{r\sqrt{\Delta_r}\,[ra^4y+(yr^3+r+2)a^2+r^3]}.
\end{eqnarray}
These functions are well defined in the region where $\Delta_r(r;a^2,y)>0$, i.e., at all radii of the stationary regions except the radii of horizons where the functions diverge as well as in the ring singularity given by the equation $\rho^2=0$. Note that the zero points of the functions also enable us to discuss changes of the orientation of the forces.
It is immediately clear from the relation~(\ref{63}) that the Coriolis force can not vanish for $v\neq0$. Therefore we will restrict our attention only to the gravitational and centrifugal forces.        

\subsection{Gravitational force} 

It follows from the relation~(\ref{61}) that for given values of the rotational and cosmological parameters $a^2$ and $y$, the radii 
of circular orbits where the gravitational force vanishes are given by solutions of the equation 
\begin{eqnarray}
\label{66}
y=y_{G\pm}(r;a^2)\equiv\frac{-r^4-ra^2(r+4)\pm\sqrt{r(a^2+r^2)[r^5+r^2a^2(r+12)+4a^4]}}{2r^2a^2(a^2+r^2)}.
\end{eqnarray}
Because of the reality condition of the forces $\Delta_r(r;a^2,y)>0$ and the repulsive cosmological constant, the solutions 
are restricted by the relation
\begin{eqnarray}
\label{64}
0<y_G(r;a^2)<y_h(r;a^2)
\end{eqnarray} 
and the plus sign in the function~(\ref{66}) is only relevant, i.e., $y_{G+}(r;a^2)\equiv y_G(r;a^2)$.
In order to obtain a number of the solutions of~(\ref{66}) and~(\ref{64}) in dependence on the parameters $a^2$ and $y$, i.e., 
to classify the KdS spacetimes according to the number of circular orbits where $\mathcal{G}(r;a^2,y)=0$, we have 
to determine the properties of the functions $y_G(r;a^2)$ and $y_h(r;a^2)$.

The asymptotic behavior of the function $y_G(r;a^2)$ is given by $y_G(r\rightarrow\infty;a^2)\rightarrow +0$,  
$y_G(r\rightarrow 0;a^2)\rightarrow\infty$, whereas $y_G(r\rightarrow 0;a^2)<y_h(r\rightarrow0;a^2)$ and
$y_G(r\rightarrow \infty;a^2)<y_h(r\rightarrow\infty;a^2)$. The common points of $y_G(r;a^2)$ and $y_h(r;a^2)$ 
are given by the relation 
\begin{eqnarray}
\label{74}
a^2=a^2_{Gh}(r)\equiv\frac{1}{2}(-2r^2+\sqrt{8r+1}r+r).
\end{eqnarray}
The maximum of the function $a^2_{Gh}(r)$ is located at $r\doteq1.616$ and takes the value $a^2_{Gh,max}\doteq1.212$. Since 
$a^2_{Gh}(r)$ and $a^2_{he}(r)$ are identical, the common points of $y_G(r;a^2)$ and $y_h(r;a^2)$ coincide with the extrema 
of $y_h(r;a^2)$. The zero points of $y_G(r;a^2)$ are given by the relation
\begin{eqnarray}
\label{79}
r^4+2r^2a^2-4ra^2+a^4=0,
\end{eqnarray}
which we consider as the implicit form of the function $a^2_{G0}(r)$. 
The maximum of this function takes the value $a^2_{G0,max}\doteq 1.688$ and is located at $r\doteq0.750$. 
The common point of $a^2_{G0}(r)$ and $a^2_{Gh}(r)$ is located at $r=1$ and takes the value $a^2=1$.
The local extrema of $y_G(r;a^2)$ are determined (due to the condition $\partial_r\: y_G(r;a^2)=0$) by the equation
\begin{eqnarray}
\label{80}
6r^2+2a^2-\frac{(r^2+a^2)(9r^4+8a^2r^2+3a^4)}{\sqrt{r(r^2+a^2)[r^5+r^2a^2(r+12)+4a^2]}}=0,\end{eqnarray}
which we consider as the implicit form of the function $a^2_{Ge}(r)$ with the maximum $a^2_{Ge,max}\doteq 2.441$ at $r\doteq1.357$.

\begin{figure}[pb]
\centerline{\psfig{file=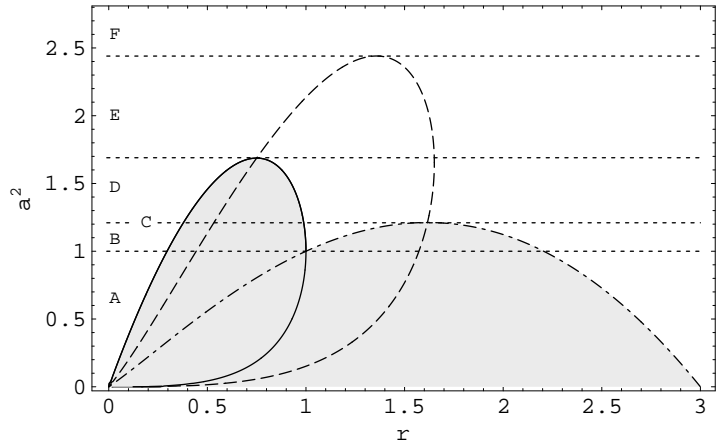,width=0.75\hsize}}
\vspace*{8pt}
\caption{Characteristic functions $a^2_{Gh}(r)$ (dashed-dotted) governing the common points of the functions $y_G(r;a^2)$ 
and $y_h(r;a^2)$ coalescent with the extrema of $y_h(r;a^2)$; $a^2_{G0}(r)$ (solid) governing the zeros of $y_G(r;a^2)$; 
$a^2_{Ge}(r)$ (dashed) governing the loci of the extrema of $y_G(r;a^2)$. 
Inside the gray region, there is  $y_G(r;a^2)<0$ or $y_G(r;a^2)>y_h(r;a^2)$, i.e, the circular orbits where $\mathcal{G}(r;a^2,y)=0$ 
cannot occur there. For given $a^2$, the common points of $y_G(r;a^2)$ and $y_h(r;a^2)$, the zeros, and the extrema of $y_G(r;a^2)$, 
respectively, are determined by the solutions of $a^2=a^2_{Gh}(r)$, $a^2=a^2_{G0}(r)$, and $a^2=a^2_{Ge}(r)$, respectively. In the parameter line $(a^2)$, the extrema and common points of the 
characteristic functions separate six regions A-F corresponding to different behavior of $y_G(r;a^2)$ and $y_h(r;a^2)$ (see Fig.~\ref{Fig:3}).}
\label{Fig:2}
\end{figure}

\begin{figure}[pb]
\centerline{\psfig{file=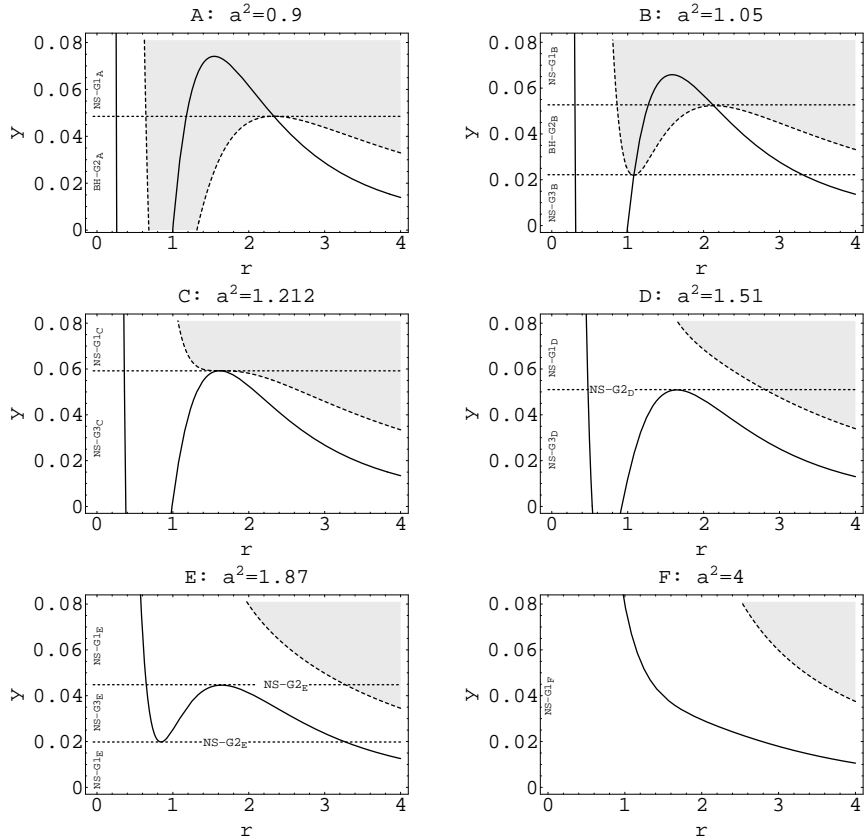,width=\hsize}}
\vspace*{8pt}
\caption{Function $y_G(r,a^2)$ (solid) determining the radii of circular orbits where $\mathcal{G}(r;a^2,y)=0$; $y_h(r,a^2)$ 
(dashed) determining the location of event horizons and limiting the dynamic region where $y>y_h(r;a^2)$ (gray). There are 
examples of qualitatively different types of behavior of $y_G(r;a^2)$ and $y_h(r;a^2)$.  For given $y$ and $a^2$, the orbits where $\mathcal{G}(r;a^2,y)=0$ are determined by  
solutions of $y=y_G(r;a^2)$ and restricted by the conditions $0<y_G(r;a^2)<y_h(r;a^2)$ (i.e., the negative or zero parts of $y_G(r;a^2)$ and the parts in the gray region or on its border are irrelevant). The horizons are determined by  
solutions of $y=y_h(r;a^2)$. In the parameter line ($y$) of the spacetimes, the extrema of $y_G(r;a^2)$ and $y_h(r;a^2)$ 
separate the regions corresponding to the KdS spacetimes BH-G2, NS-G1, NS-G2, and NS-G3 differing in the number of 
circular orbits where $\mathcal{G}(r;a^2,y)=0$ for given values of the rotational parameter $a^2$.}
\label{Fig:3}
\end{figure}

All the characteristic functions $a^2_{Gh}(r)$, $a^2_{G0}(r)$, and $a^2_{Ge}(r)$ are illustrated in Fig.~\ref{Fig:2}.
These functions enable us to understand the different behavior of the functions $y_G(r;a^2)$ and $y_h(r;a^2)$ (see Fig.~\ref{Fig:3}). For $a^2<1.212$, 
$y_G(r;a^2)$ has two common points with $y_h(r;a^2)$ determined by the equations~(\ref{66}) and~(\ref{74}). For $a^2=1.212$, these 
points coincide and for $a^2>1.212$, there are no such points. For $a^2<1.688$, $y_G(r;a^2)$ has two zero points determined 
by the Eq.~(\ref{79}). For $a^2=1.688$, these points coincide and for $a^2>1.688$, there are no such points. For $a^2<2.441$, 
$y_G(r;a^2)$ has two local extrema $y_{G,min}(a^2)$ and $y_{G,max}(a^2)$ determined by the equations~(\ref{66}) and~(\ref{80}). For $a^2=2.441$, these extrema coincide and for $a^2>2.441$, there are no extrema of $y_G(r;a^2)$.

The radii of circular orbits where the gravitational force has its local extrema are determined by solutions of the equation
$\partial_r(\mathcal{G})=0$, which we consider as the implicit form of the function $y_{G,ext}(r;a^2)$ (we do not give this 
expression here due to its length). The gravitational force changes its orientation at points of its vanishing if and only if 
it has no local extrema there. Thus common points of the functions $y_{G,ext}(r;a^2)$ and $y_{G}(r;a^2)$ determine the circular 
orbits where the force vanishes but does not change its orientation. These common points are determined by the equation identical 
with~(\ref{80}), i.e., they are governed by the function $a^2_{Ge}(r)$ and coincide with the extrema of the function $y_G(r;a^2)$.

The classification of the KdS spacetimes according to the number of circular orbits where $\mathcal{G}(r;a^2,y)=0$ 
can be now given in the following way. We can distinguish six different types of behavior of $y_G(r;a^2)$ and $y_h(r;a^2)$.

\paragraph{Type A:\quad $0<a^2\leq 1$}(Fig.~\ref{Fig:3}A)\\
The functions $y_G(r;a^2)$ and $y_h(r;a^2)$ have both two extrema; $y_{G,max}(a^2)>y_{h,max}(a^2)>0$, 
$y_{G,min}(a^2)<y_{h,min}(a^2)\leq0$. Spacetimes with $y\geq y_{h,max}(a^2)$ (NS-G1{\scriptsize A}) contain 
one circular orbit where the gravitational force vanishes and changes its orientation.  Spacetimes with $0<y<y_{h,max}(a^2)$ 
(BH-G2{\scriptsize A}) contain two such orbits, whereas the inner one occurs in the inner BH region and the outer one occurs 
in the outer BH region. Note that here, and henceforth, the values of $y_G(r;a^2)\geq y_{h}(r;a^2)$ as well as $y_G(r;a^2) \leq 0$ are irrelevant because they do not satisfy the condition~(\ref{64}).
\paragraph{Type B:\quad $1<a^2<1.212$}(Fig.~\ref{Fig:3}B)\\
The functions $y_G(r;a^2)$ and $y_h(r;a^2)$ have both two extrema;  $y_{G,min}(a^2)<0$, 
$0<y_{h,min}(a^2)<y_{h,max}(a^2)<y_{G,max}(a^2)$. Spacetimes with  $y\geq y_{h,max}(a^2)$  (NS-G1{\scriptsize B}) contain one 
circular orbit where the gravitational force vanishes and changes its orientation, while spacetimes with $0<y<y_{h,min}(a^2)$ 
(NS-G3{\scriptsize B}) contain three such orbits. Spacetimes with $y_{h,min}(a^2)\leq y<y_{h,max}(a^2)$ (BH-G2{\scriptsize B}) 
contain two such orbits. The inner one occurs in the inner BH region and the outer one occurs in the outer BH region.
\paragraph{Type C:\quad $a^2=1.212$}(Fig.~\ref{Fig:3}C)\\
The function $y_G(r;a^2)$ has two extrema and $y_h(r;a^2)$ has no extrema; $y_{G,min}<0$, $y_{G,max}\in y_h(r;a^2)$. 
Spacetimes with  $y \geq y_{G,max}(a^2)$ (NS-G1{\scriptsize C}) contain one circular orbit where the gravitational force vanishes 
and changes its orientation, while spacetimes with $0<y<y_{G,max}(a^2)$ (NS-G3{\scriptsize C}) contain three such orbits. 
\paragraph{Type D:\quad $1.212<a^2\leq 1.688$}(Fig.~\ref{Fig:3}D)\\
The function $y_G(r;a^2)$ has two extrema and $y_h(r;a^2)$ has no extrema; $y_{G,min}\leq 0$, $y_{G,max}>0$.
Spacetimes with  $y>y_{G,max}(a^2)$ (NS-G1{\scriptsize D}) contain one circular orbit where the gravitational force vanishes 
and changes its orientation, while spacetimes with $0<y<y_{G,max}(a^2)$ \mbox{(NS-G3{\scriptsize D})} contain three such orbits. 
Spacetimes with $y=y_{G,max}(a^2)$ (NS-G2{\scriptsize D}) contain two orbits where the gravitational force vanishes, but 
the orientation is changed only on the inner orbit.
\paragraph{Type E:\quad $1.688<a^2<2.441$}(Fig.~\ref{Fig:3}E)\\
The function $y_G(r;a^2)$ has two extrema and $y_h(r;a^2)$ has no extrema; $0<y_{G,min}(a^2)<y_{G,max}(a^2)$.
Spacetimes with  $y>y_{G,max}(a^2)$ or $0<y<y_{G,min}(a^2)$ (NS-G1{\scriptsize E}) contain one circular orbit where the 
gravitational force vanishes and changes its orientation, while spacetimes with $y_{G,min}(a^2)<y<y_{G,max}(a^2)$ 
(NS-G3{\scriptsize E}) contain three such orbits. Spacetimes with $y=y_{G,min}(a^2)$ or $y=y_{G,max}(a^2)$ 
(NS-G2{\scriptsize E}) contain two orbits where the gravitational force vanishes, but the orientation is changed only on 
one of these orbits.  
\paragraph{Type F:\quad $2.441\leq a^2<\infty$}(Fig.~\ref{Fig:3}F)\\
The functions $y_G(r;a^2)$ and $y_h(r;a^2)$ have no extrema. All the spacetimes with $0<y<\infty$ (NS-G1{\scriptsize F}) 
contain only one circular orbit where the gravitational force vanishes and changes its orientation.\\

In the parameter plane $(a^2,y)$ of the KdS spacetimes, the functions  $y_{G,min}(a^2)$, $y_{G,max}(a^2)$, 
$y_{h,max}(a^2)$, and $y_{h,min}(a^2)$ separate four regions corresponding to one class of the KdS black-hole spacetimes BH-G2\\ 
\begin{center}
\begin{tabular}{cccll}
\hline
\bf{Class}&\bf{Subclass}&\bf{Orbits}&\bf{Interval of} $\mathbf{a^2}$&\bf{Interval of} $\mathbf{y}$\\
\hline
BH-G2&BH-G2\scriptsize{A}&2&$(0,1\rangle$&$(0,y_{h,max}(a^2))$\\
&BH-G2\scriptsize{B}&2&$(1,1.212)$&$\langle y_{h,min}(a^2),y_{h,max}(a^2))$\\
\hline
\end{tabular}
\end{center}
\quad\\
and to three classes of the KdS naked-singularity spacetimes NS-G1, NS-G2, and NS-G3 differing in the number of 
circular orbits where $\mathcal{G}(r;a^2,y)=0$. 
\begin{center}
\begin{tabular}{cccll}
\hline
\bf{Class}&\bf{Subclass}&\bf{Orbits}&\bf{Interval of} $\mathbf{a^2}$&\bf{Interval of} $\mathbf{y}$\\
\hline
NS-G1&NS-G1\scriptsize{A}&1&$(0,1\rangle$&$\langle y_{h,max}(a^2),\infty)$\\ 
&NS-G1\scriptsize{B}&1&$(1,1.212)$&$\langle y_{h,max}(a^2),\infty)$\\
&NS-G1\scriptsize{C}&1&$a^2=1.212$&$\langle y_{G,max}(a^2),\infty)$\\
&NS-G1\scriptsize{D}&1&$(1.212,1.688\rangle$&$(y_{G,max}(a^2),\infty)$\\
&NS-G1\scriptsize{E}&1&$(1.688,2.441)$&$(0,y_{G,min}(a^2))\cup (y_{G,max}(a^2),\infty)$\\
&NS-G1\scriptsize{F}&1&$\langle 2.441,\infty)$&$(0,\infty)$\\  
NS-G3&NS-G3\scriptsize{B}&3&$(1,1.212)$&$(0,y_{h,min}(a^2))$\\ 
&NS-G3\scriptsize{C}&3&$a^2=1.212$&$(0,y_{G,max}(a^2))$\\ 
&NS-G3\scriptsize{D}&3&$(1.212,1.688\rangle$&$(0,y_{G,max}(a^2))$\\ 
&NS-G3\scriptsize{E}&3&$(1.688,2.441)$&$(y_{G,min}(a^2),y_{G,max}(a^2))$\\
NS-G2&NS-G2\scriptsize{D}&2&$(1.212,1.688\rangle$&$y=y_{G,max}(a^2)$\\
&NS-G2\scriptsize{E}&2&$(1.688,2.441)$&$y=y_{G,min}(a^2) \cup y_{G,max}(a^2)$\\  
\hline
\end{tabular}
\end{center}

\begin{figure}[pb]
\centerline{\psfig{file=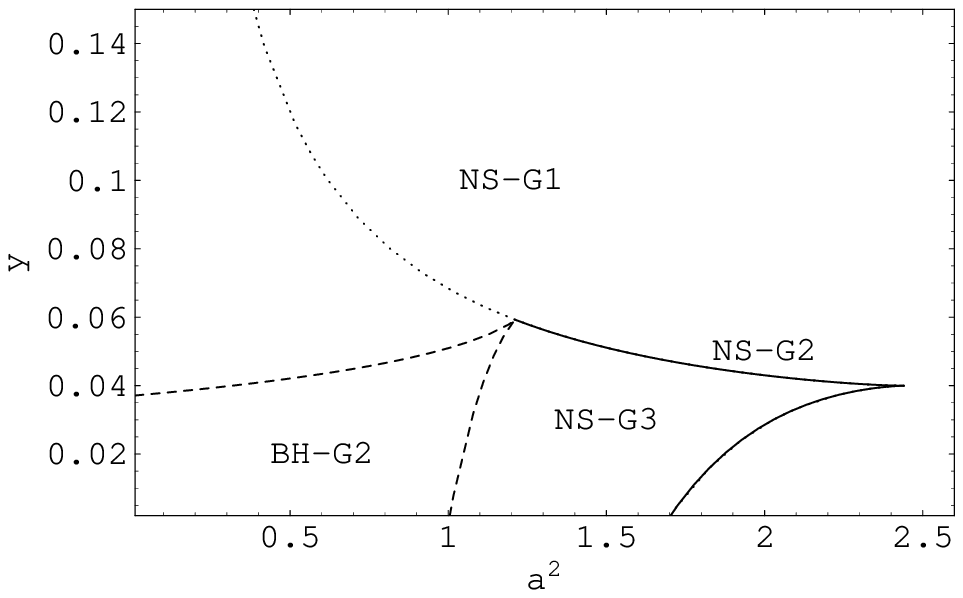,width=0.7\hsize}}
\vspace*{8pt}
\caption{The classification of the KdS spacetimes according to the number of circular orbits where 
$\mathcal{G}(r;a^2,y)=0$. The functions $y_{h,max}(a^2)$ (dashed-upper) and $y_{h,min}(a^2)$ (dashed-lower) determining 
the extrema of $y_h(r;a^2)$ separate regions corresponding to the black-hole spacetimes BH-G2 with two circular orbits where $\mathcal{G}(r;a^2,y)=0$
and to the naked-singularity spacetimes. For $a^2>1.212$, the functions $y_{G,max}(a^2)$  
(solid-upper) and $y_{G,min}(a^2)$ (solid-lower) determining the extrema of $y_G(r;a^2)$ 
determine the spacetimes NS-G2 with two circular orbits and separate the regions corresponding to the \mbox{NS-G1} and \mbox{NS-G3} 
naked-singularity spacetimes with one and three circular orbits where $\mathcal{G}(r;a^2,y)=0$. For \mbox{$a^2\leq 1.212$}, 
the function $y_{G,max}(a^2)$ (dotted) does not influence the classification as well as the negative parts of the 
functions $y_{h,min}(a^2)$ and $y_{G,min}(a^2)$ which are not illustrated here.}
\label{Fig:4}
\end{figure}

\begin{figure}[pb]
\centerline{\psfig{file=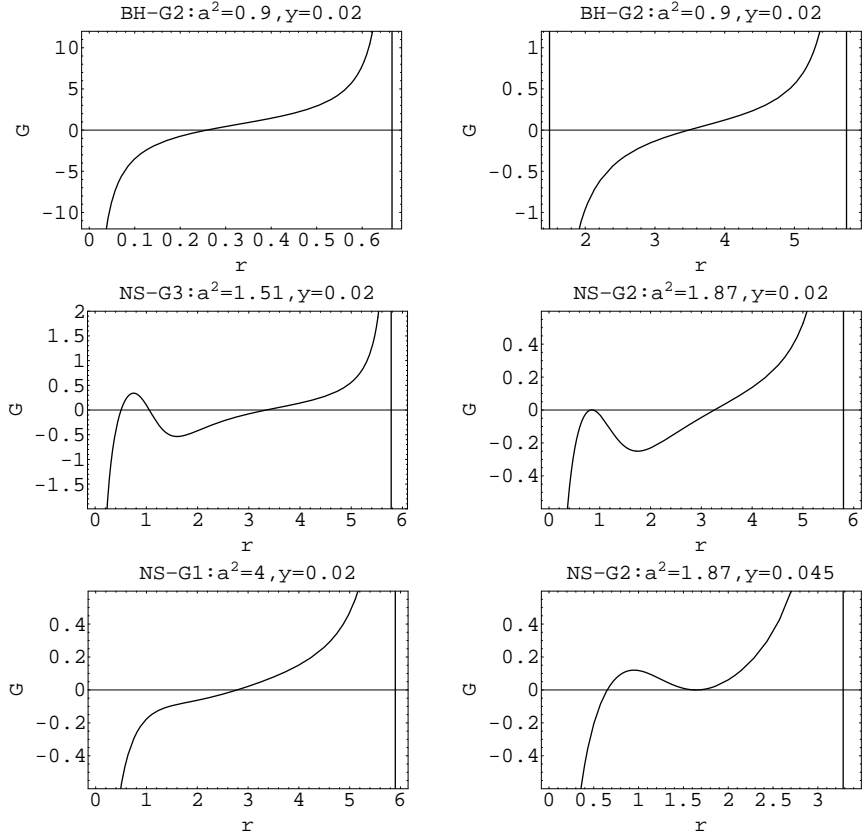,width=1\hsize}}
\vspace*{8pt}
\caption{The mass independent part of the gravitational force. There are examples of qualitatively different types of behavior of $\mathcal{G}(r;a^2,y)$ corresponding to the classification of the KdS spacetimes according to the number of circular orbits where $\mathcal{G}(r;a^2,y)=0$. The first type of behavior corresponding to the class BH-G2 is illustrated in the inner and outer stationary black-hole regions separately. Positive (negative) parts of $\mathcal{G}(r;a^2,y)$ determine repulsive (attractive) force oriented outwards (towards) the singularity at $r=0$. The zero points of the function determine the radii of circular orbits where the gravitational force vanishes. The vertical lines denote the asymptotic behavior of the function at event horizons.}
\label{Fig:5}
\end{figure}

In Fig.~\ref{Fig:4} we give the behavior of the functions  $y_{G,min}(a^2)$, $y_{G,max}(a^2)$, $y_{h,max}(a^2)$, and 
$y_{h,min}(a^2)$. Qualitatively different types of behavior of the function $\mathcal{G}(r;a^2,y)$ corresponding to the given classification are illustrated 
in Fig.~\ref{Fig:5}. 
 
\subsubsection*{Schwarzschild-de~Sitter and Kerr cases}
\begin{figure}[pb]
\centerline{\psfig{file=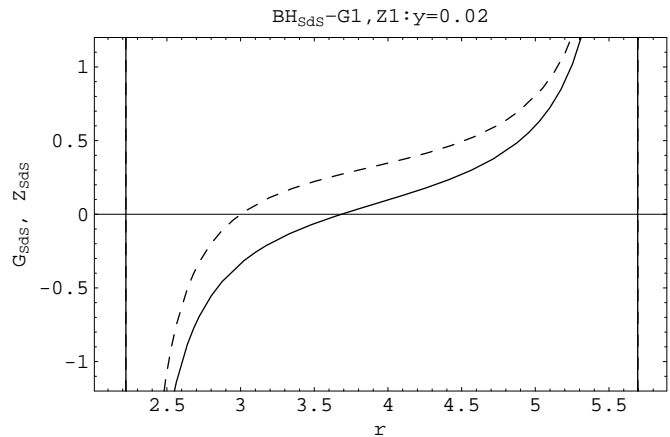,width=0.7\hsize}}
\vspace*{8pt}
\caption{The mass and velocity independent parts of the gravitational (solid) and centrifugal (dashed) forces. There are examples of the only type of behavior of $\mathcal{G}_{SdS}(r;y)$ and $\mathcal{Z}_{SdS}(r;y)$ corresponding to the classes BH{\scriptsize{SdS}}-G1 and BH{\scriptsize{SdS}}-Z1. Positive (negative) parts of $\mathcal{G}_{SdS}(r;y)$ and $\mathcal{Z}_{SdS}(r;y)$ determine repulsive (attractive) forces oriented outwards (towards) the singularity at $r=0$. The zero points of the functions determine the radii of circular orbits where the forces vanish independently of the velocity. The vertical lines denote the asymptotic behavior of the functions at event horizons.}
\label{Fig:51}
\end{figure}

In the SdS spacetimes \cite{Stu:1990:BULAI:} ($a^2=0$, $y>0$), we obtain from the relation~(\ref{61}) that the mass independent 
part of the gravitational force is given by the relation 
\begin{eqnarray}
\label{82}
\mathcal{G}_{SdS}(r;y)=\frac{r^3y-1}{\Delta_r},\end{eqnarray}
where $\Delta_r=r^2-2r-yr^4$. The gravitational force is well defined in the stationary regions where $\Delta_r>0$, i.e., at 
all radii between the black-hole and cosmological event horizons given by solutions of the equation 
\begin{eqnarray}
\label{83}
y=y_{hSdS}(r)\equiv \frac{r-2}{r^3}.
\end{eqnarray}
Both horizons coincide in the case of $y=1/27$ at $r=3$ and for $y>1/27$ there are no horizons, i.e., all the spacetime is dynamic.   
The function $\mathcal{G}_{SdS}(r;y)$ diverges at the radii of horizons and vanishes only at the static radius $r=y^{-1/3}$  where the gravitational attraction of the black hole is balanced by the cosmological repulsion \cite{Stu-Hle:1999:PHYSR4:}. In the case of $y=1/27$ the static radius coincide with the horizons. Thus there is only one class of the stationary SdS spacetimes BH{\scriptsize{SdS}}-G1 ($0<y<1/27$) with one circular orbit at the static radius  where the gravitational force vanishes and changes its orientation (see Fig.~\ref{Fig:51}). 

The circular orbit with the static radius has an important meaning even in the KdS spacetimes. It is the outer limit for existence of circular geodesics where 
two families of the geodesic circular stationary motion coalesce \cite{Stu-Sla:2004:PHYSR4:}. 
According to the equation~(\ref{41a}), the radii of circular geodesics satisfy the condition  
\begin{eqnarray}
\label{84}
\mathcal{G}(r;a^2,y)+(\gamma v)^2\,\mathcal{Z}(r;a^2,y)+\gamma^2 v\,\mathcal{C}(r;a^2,y)=0.
\end{eqnarray}
It can be shown that the relation $r\leq y^{-1/3}$ is the condition for the velocity $v$ to be real. Thus it is also the 
necessary condition for the stationary circular geodesic motion in the equatorial plane of the KdS spacetimes.

In the Kerr spacetimes \cite{Stu-Hle:1999:ACTPS2:} ($a^2>0$, $y=0$), we obtain from the relation~(\ref{61}) the mass independent part of the gravitational 
force in the form 
\begin{eqnarray}
\label{85}
\mathcal{G}_{K}(r;a^2)=\frac{-r^4-2r^2a^2+4ra^2-a^4}{r\Delta_r[(r+2)a^2+r^3]},
\end{eqnarray}
where $\Delta_r=r^2-2r+a^2$. In the Kerr spacetimes, the gravitational force is well defined in the stationary 
region where $\Delta_r>0$, i.e., it is not defined in the dynamic region between the inner and outer black-hole event horizons 
determined by solutions of the equation 
\begin{eqnarray}
\label{86}
a^2=a^2_{hK}(r)\equiv r(2-r).
\end{eqnarray}
The maximum of the function $a^2_{hK}(r)$ is located at $r=1$ and takes the value $a^2_{hK,max}=1$. For $a^2<1$, there are 
black-hole spacetimes containing two horizons, which  coincide for $a^2=1$. The naked-singularity spacetimes with no horizon 
exist for $a^2>1$. 
It follows from the relation~(\ref{85}) that for given values of the rotational parameter $a^2$, the gravitational force 
vanishes on circular orbits with the radii determined by solutions of the equation
\begin{eqnarray}
\label{87}
r^4+2r^2a^2-4ra^2+a^4=0,
\end{eqnarray}
which we consider as the implicit form of the function $a^2_{GK}(r)$, and restricted by the condition 
\begin{eqnarray}
\label{88}
a^2_{GK}(r)>a^2_{hK}(r).
\end{eqnarray}
The function $a^2_{GK}(r)$ clearly coincides with the function $a^2_{G0}(r)$. Its maximum is located at $r\doteq0.750$ and takes the 
value $a^2_{GK,max}\doteq1.688$. For given values of the rotational parameter $a^2$, the local extrema of the function 
$\mathcal{G}_{K}(r;a^2)$ are given by solutions of the equation 
\begin{eqnarray}
\label{89}
a^8(1+r)+4a^6r[r(r-1)-1]+2a^4r^2[r^2(3r-10)+4]+\nonumber\\
4a^2r^5[(r(r-4)+5]+r^8(r-1)=0,
\end{eqnarray}
which we consider as the implicit form of the function $a^2_{GeK}(r)$. The radii of circular orbits where the gravitational 
force vanishes and does not change its orientation are determined by common points of $a^2_{GK}(r)$ and $a^2_{GeK}(r)$. The 
only such point satisfying the condition~(\ref{88}) corresponds to the maximum of  $a^2_{GK}(r)$.   

\begin{figure}[pb]
\centerline{\psfig{file=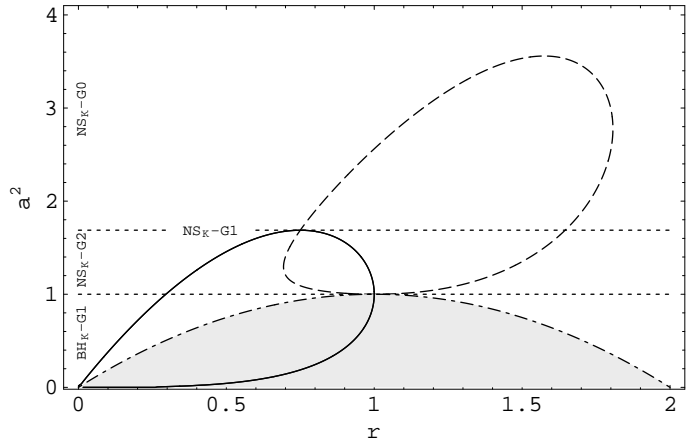,width=0.75\hsize}}
\vspace*{8pt}
\caption{Characteristic functions $a^2_{hK}(r)$ (dashed-dotted) determining the loci of event horizons and limiting the 
dynamic region where $a^2<a^2_{hK}(r)$ (gray); $a^2_{GK}(r)$ (solid) determining the circular orbits where 
$\mathcal{G}_{K}(r;a^2)=0$; $a^2_{GeK}(r)$ (dashed) determining the extrema of $\mathcal{G}_{K}(r;a^2)$.  
For given $a^2$, the loci of horizons are determined by the solutions of $a^2=a^2_{hK}(r)$. The circular orbits where 
$\mathcal{G}_{K}(r;a^2)=0$ and the extrema of $\mathcal{G}_{K}(r;a^2)$, respectively, are determined by the solutions of 
$a^2=a^2_{GK}(r)$ and $a^2=a^2_{GeK}(r)$, respectively, and restricted by the conditions $a^2_{GK}(r)>a^2_{hK}(r)$ and $a^2_{GeK}(r)>a^2_{hK}(r)$ (i.e., the parts of $a^2_{GK}(r)$ and $a^2_{GeK}(r)$ in the gray region or on its border are irrelevant).  
In the parameter line ($a^2$), the extrema of the characteristic functions separate regions corresponding to one 
class of the Kerr black-hole spacetimes BH{\scriptsize{K}}-G1 with one circular orbits and to three classes of the 
Kerr naked-singularity spacetimes NS{\scriptsize{K}}-G2, NS{\scriptsize{K}}-G1, and \mbox{NS{\scriptsize{K}}-G0} with (subsequently) 
two, one, and none circular orbits where $\mathcal{G}_{K}(r;a^2)=0$.}
\label{Fig:6}
\end{figure}

\begin{figure}[pb]
\centerline{\psfig{file=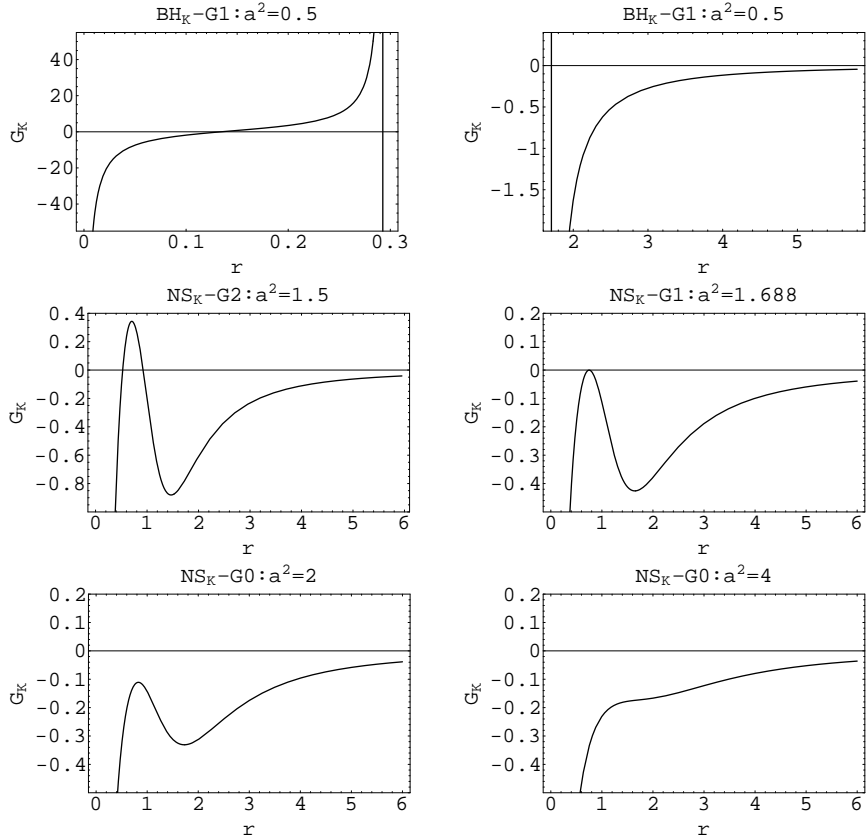,width=\hsize}}
\vspace*{8pt}
\caption{The mass independent part of the gravitational force. There are examples of qualitatively different types of behavior of $\mathcal{G}_K(r;a^2)$ corresponding to the classification of the Kerr spacetimes according to the number of circular orbits where $\mathcal{G}_K(r;a^2)=0$. The first type of behavior corresponding to the class BH{\scriptsize{K}}-G1  is illustrated in the inner and outer stationary black-hole regions separately. Positive (negative) parts of $\mathcal{G}_K(r;a^2)$ determine repulsive (attractive) force oriented outwards (towards) the singularity at $r=0$. The zero points of the function determine the radii of circular orbits where the gravitational force vanishes. The vertical lines denote the asymptotic behavior of the function at event horizons.}
\label{Fig:7}
\end{figure}

All the characteristic functions $a^2_{hK}(r)$, $a^2_{GK}(r)$, and $a^2_{GeK}(r)$ are illustrated in Fig.~\ref{Fig:6}.
In the parameter line ($a^2$) of the Kerr spacetimes, the extrema of the characteristic functions separate regions corresponding to one class of the Kerr black-hole spacetimes BH{\scriptsize{K}}-G1 and to three classes of naked-singularity spacetimes NS{\scriptsize{K}}-G0, NS{\scriptsize{K}}-G1, and NS{\scriptsize{K}}-G2 differing in the number of circular orbits where $\mathcal{G}_{K}(r;a^2)=0$. 
\paragraph{Class BH{\scriptsize{K}}-G1:\quad $0<a^2\leq 1$}\quad\\
In the inner stationary region of the spacetime, there is one orbit where the gravitational force vanishes and changes its 
orientation. In the outer stationary region, there is no such orbit.
\paragraph{Class NS{\scriptsize{K}}-G2:\quad $1<a^2<1.688$}\quad\\
There are two orbits where the gravitational force vanishes and changes its orientation. Note that the inversion of orientation of the 
gravitational force in the field of Kerr naked singularities was first noticed by de Felice \cite{deFel:1974:ASTRA:}.
\paragraph{Class NS{\scriptsize{K}}-G1:\quad $a^2=1.688$}\quad\\
There is one orbit where the gravitational force vanishes but does not change its orientation.
\paragraph{Class NS{\scriptsize{K}}-G0:\quad $a^2>1.688$}\quad\\
There is no orbit where the gravitational force vanishes. 
\\\\
Qualitatively different types of behavior of the function $\mathcal{G}_{K}(r;a^2)$    corresponding to the given classification are illustrated in \mbox{Fig.~\ref{Fig:7}.}

\subsection{Centrifugal force}

It follows from the relation~(\ref{62}) that for given values of the rotational and cosmological parameters $a^2$ and $y$, the 
radii of circular orbits where the centrifugal force vanishes independently of the velocity are given by solutions of the equation 
\begin{eqnarray}
\label{66z}
y=y_{Z\pm}(r;a^2)\equiv\{2a^2r^2(a^2+r^2)\}^{-1}\{-(2r+5)ra^2-r^3(2r+3)\pm\nonumber\\
\sqrt{r(a^2+3r^2)[8a^4+ra^2(16r+1)+r^3(8r+3)]}\}.
\end{eqnarray}
Because of the reality condition of the forces $\Delta_r(r;a^2,y)>0$ and the repulsive cosmological constant, the solutions 
are restricted by the relation
\begin{eqnarray}
\label{64z}
0<y_Z(r;a^2)<y_h(r;a^2)
\end{eqnarray} 
and the plus sign in the function~(\ref{66z}) is only relevant, i.e., $y_{Z+}(r;a^2)\equiv y_Z(r;a^2)$.
In order to obtain a number of the solutions of~(\ref{66z}) and~(\ref{64z}) in dependence on the parameters $a^2$ and $y$, i.e., 
to classify the KdS spacetimes according to the number of circular orbits where $\mathcal{Z}(r;a^2,y)=0$, we have to 
determine the properties of the functions $y_Z(r;a^2)$ and $y_h(r;a^2)$. 

The asymptotic behavior of the function $y_Z(r;a^2)$ 
is given by $y_Z(r\rightarrow\infty;a^2)\rightarrow -\infty$, $y_Z(r\rightarrow 0;a^2)\rightarrow\infty$, whereas  
$y_Z(r\rightarrow 0;a^2)<y_h(r\rightarrow0;a^2)$ and $y_Z(r\rightarrow \infty;a^2)<y_h(r\rightarrow\infty;a^2)$.
The common points of $y_Z(r;a^2)$ and $y_h(r;a^2)$ are given by the relation
\begin{eqnarray}
\label{74z}
a^2=a^2_{Zh}(r)\equiv\frac{1}{2}(-2r^2+\sqrt{8r+1}r+r).
\end{eqnarray} 
The function $a^2_{Zh}(r)$ is identical with the function $a^2_{Gh}(r)$ and its maximum is located at $r\doteq1.616$ and takes 
the value $a^2_{Zh,max}\doteq1.212$. Since $a^2_{Zh}(r)$ and  $a^2_{he}(r)$ are also identical, the common points of 
$y_Z(r;a^2)$ and $y_h(r;a^2)$ coincide with the extrema of $y_h(r;a^2)$. The zero points of $y_Z(r;a^2)$ are given by the relation
\begin{eqnarray}
\label{79z}
r^4(r-3)+ra^2[r(r-3)+6]-2a^4=0,
\end{eqnarray}
which we consider as the implicit form of the function $a^2_{Z0}(r)$. 
The maximum of this function takes the value $a^2_{Z0,max}\doteq 1.367$ and is located at $r\doteq0.812$. 
The common point of $a^2_{Z0}(r)$ and $a^2_{Zh}(r)$ is located at $r=1$ and takes the value $a^2=1$.
The local extrema of $y_Z(r;a^2)$ are determined (due to the condition $\partial_r\: y_Z(r;a^2)=0$) by the equation
\begin{eqnarray}
\label{80z}
\{\sqrt{r(a^2+3r^2)[8a^4+r(16r+1)a^2+r^3(8r+3)]}\}^{-1}\times\nonumber\\
\,[12a^8+r(48r+1)a^6+3r^3(24r+1)a^4+3r^5(16r+1)a^2+\nonumber\\
3r^7(4r+3)]-5a^4-12r^2a^2-3r^4=0,
\end{eqnarray}
which we consider as the implicit form of the function $a^2_{Ze}(r)$ with the maximum $a^2_{Ze,max}\doteq 1.813$ at $r\doteq1.329$.

\begin{figure}[pb]
\centerline{\psfig{file=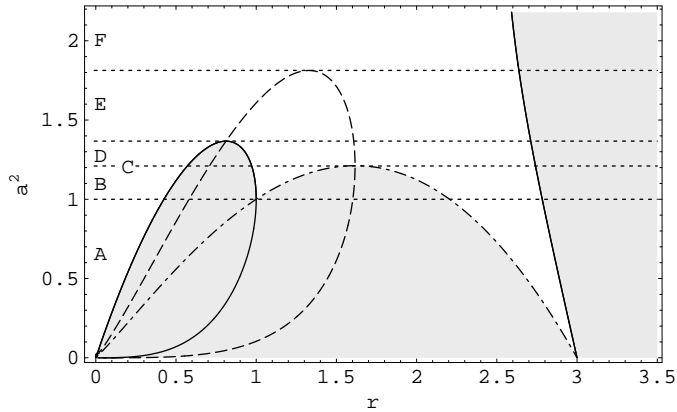,width=0.7\hsize}}
\vspace*{8pt}
\caption{Characteristic functions $a^2_{Zh}(r)$ (dashed-dotted) governing the common points of the functions $y_Z(r;a^2)$ 
and $y_h(r;a^2)$ coalescent with the extrema of $y_h(r;a^2)$; $a^2_{Z0}(r)$ (solid) governing the zeros of $y_Z(r;a^2)$; 
$a^2_{Ze}(r)$ (dashed) governing the loci of extrema of $y_Z(r;a^2)$. 
Inside the gray region, there is  $y_Z(r;a^2)<0$ or $y_Z(r;a^2)>y_h(r;a^2)$, i.e, the circular orbits where 
$\mathcal{Z}(r;a^2,y)=0$ cannot occur there. For given $a^2$, the common points of $y_Z(r;a^2)$ and $y_h(r;a^2)$, the zeros, 
and the extrema of $y_Z(r;a^2)$, respectively, are determined by the solutions of $a^2=a^2_{Zh}(r)$, $a^2=a^2_{Z0}(r)$,  and 
$a^2=a^2_{Ze}(r)$, respectively. In the parameter 
line $(a^2)$, the extrema and common points of the characteristic functions separate six regions A-F corresponding to 
different behavior of $y_Z(r;a^2)$ and $y_h(r;a^2)$ (see Fig.~\ref{Fig:3z}).}   
\label{Fig:2z}
\end{figure}

\begin{figure}[pb]
\centerline{\psfig{file=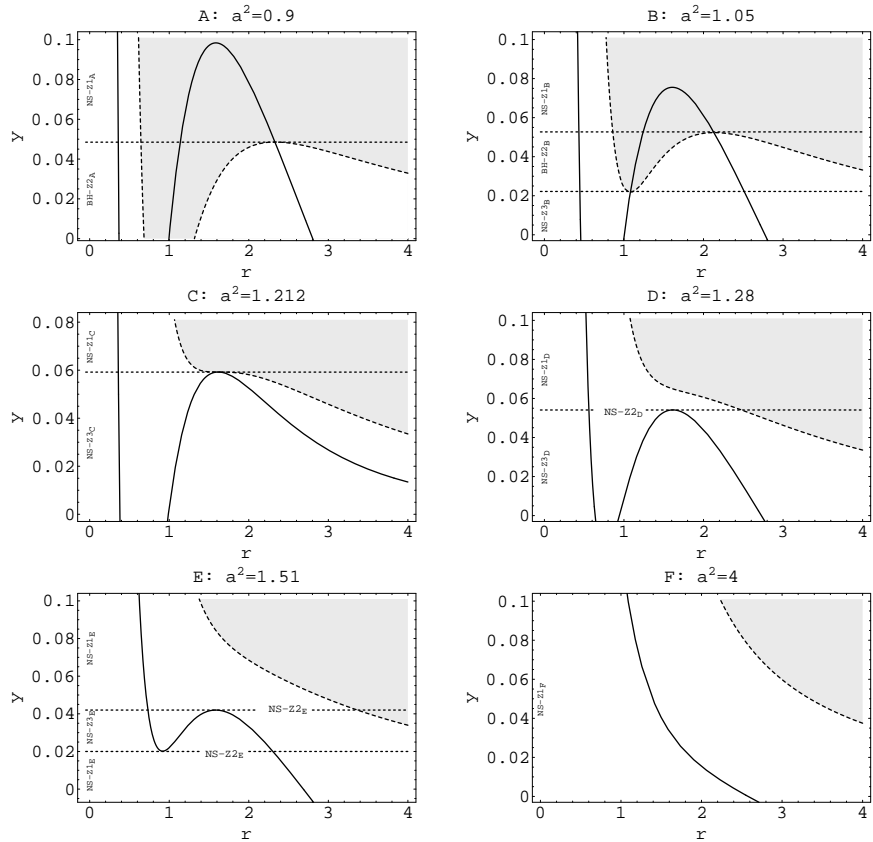,width=\hsize}}
\vspace*{8pt}
\caption{Functions $y_Z(r,a^2)$ (solid) determining the radii of circular orbits where $\mathcal{Z}(r;a^2,y)=0$; $y_h(r,a^2)$ 
(dashed) determining the location of event horizons and limiting the dynamic region where $y>y_h(r;a^2)$ (gray). There are 
examples of qualitatively different types of behavior of $y_Z(r;a^2)$ and $y_h(r;a^2)$. For given $y$ and $a^2$, the orbits where $\mathcal{Z}(r;a^2,y)=0$ are determined 
by solutions of $y=y_Z(r;a^2)$ and restricted by the conditions $0<y_Z(r;a^2)<y_h(r;a^2)$ (i.e., the negative or zero parts of $y_Z(r;a^2)$ and the parts in the gray region or on its border are irrelevant). The horizons are determined 
by solutions of $y=y_h(r;a^2)$. In the parameter line ($y$) of the spacetimes, the extrema of $y_Z(r;a^2)$ and $y_h(r;a^2)$ 
separate the regions corresponding to the KdS spacetimes BH-Z2, NS-Z1, NS-Z2, and NS-Z3 differing in the number of 
circular orbits where $\mathcal{Z}(r;a^2,y)=0$ for given values of the rotational parameter $a^2$.}
\label{Fig:3z}
\end{figure}

All the characteristic functions $a^2_{Zh}(r)$, $a^2_{Z0}(r)$, and $a^2_{Ze}(r)$ are illustrated in Fig.~\ref{Fig:2z}.
These functions enable us to understand the different behavior of the functions $y_Z(r;a^2)$ and $y_h(r;a^2)$ (see Fig.~\ref{Fig:3z}). For $a^2<1.212$, 
$y_Z(r;a^2)$ has two common points with $y_h(r;a^2)$ determined by the equations~(\ref{66z}) and~(\ref{74z}). For $a^2=1.212$, 
these points coincide and for $a^2>1.212$, there are no such points. 
For $a^2<1.367$, $y_Z(r;a^2)$ has two zero points determined by the Eq.~(\ref{79z}). For $a^2=1.367$, these points 
coincide and for $a^2>1.367$, there are no such points. For $a^2<1.813$, $y_Z(r;a^2)$ has two local extrema $y_{Z,min}(a^2)$ 
and $y_{Z,max}(a^2)$ determined by the equations~(\ref{66z}) and~(\ref{80z}). For $a^2=1.813$, these extrema coincide and for $a^2>1.813$, there are no extrema of $y_Z(r;a^2)$.

The radii of circular orbits where the centrifugal force has its local extrema are determined by solutions of the equation
$\partial_r(\mathcal{Z})=0$, which we consider as the implicit form of the function $y_{Z,ext}(r;a^2)$ (we do not give this 
expression here due to its length). The centrifugal force changes its orientation at points of its vanishing if and only if 
it has no local extrema there. Thus common points of the functions $y_{Z,ext}(r;a^2)$ and $y_{Z}(r;a^2)$ 
determine the circular orbits where the force vanishes but does not change its orientation. These common points are determined 
by the equation identical with~(\ref{80z}), i.e., they are governed by the function $a^2_{Ze}(r)$ and coincide with the extrema 
of the function $y_Z(r;a^2)$.

The classification of the KdS spacetimes according to the number of circular orbits where $\mathcal{Z}(r;a^2,y)=0$ 
can be now given in the following way. We can distinguish six different types of behavior of $y_Z(r;a^2)$ and $y_h(r;a^2)$.
\paragraph{Type A:\quad $0<a^2\leq 1$}(Fig.~\ref{Fig:3z}A)\\
The functions $y_Z(r;a^2)$ and $y_h(r;a^2)$ have both two extrema; $y_{Z,max}(a^2)>y_{h,max}(a^2)>0$, 
$y_{Z,min}(a^2)<y_{h,min}(a^2)\leq0$. Spacetimes with $y\geq y_{h,max}(a^2)$ (NS-Z1{\scriptsize A}) contain 
one circular orbit where the centrifugal force vanishes independently of the velocity and changes its orientation. Spacetimes with $0<y<y_{h,max}(a^2)$ (BH-Z2{\scriptsize A}) contain two such orbits, whereas the inner one occurs in the inner BH region and the outer one occurs in the outer BH region. Note that here, and henceforth, the values of $y_Z(r;a^2)\geq y_{h}(r;a^2)$ as well as $y_Z(r;a^2) \leq 0$ are irrelevant because they do not satisfy the condition~(\ref{64z}).
\paragraph{Type B:\quad $1<a^2<1.212$}(Fig.~\ref{Fig:3z}B)\\
The functions $y_Z(r;a^2)$ and $y_h(r;a^2)$ have both two extrema;  $y_{Z,min}(a^2)<0$, 
$0<y_{h,min}(a^2)<y_{h,max}(a^2)<y_{Z,max}(a^2)$. Spacetimes with  $y\geq y_{h,max}(a^2)$  (NS-Z1{\scriptsize B}) contain one circular orbit where the centrifugal force vanishes independently of the velocity and changes its orientation, while spacetimes with $0<y<y_{h,min}(a^2)$ (NS-Z3{\scriptsize B}) contain three such orbits. Spacetimes with $y_{h,min}(a^2)\leq y<y_{h,max}(a^2)$ (BH-Z2{\scriptsize B}) contain two such orbits. The inner one occurs in the inner BH region and the outer one occurs in the outer BH region.
\paragraph{Type C:\quad $a^2=1.212$}(Fig.~\ref{Fig:3z}C)\\
The function $y_Z(r;a^2)$ has two extrema and $y_h(r;a^2)$ has no extrema; $y_{Z,min}<0$, $y_{Z,max}\in y_h(r;a^2)$. 
Spacetimes with  $y \geq y_{Z,max}(a^2)$ (NS-Z1{\scriptsize C}) contain one circular orbit where the gravitational force vanishes 
and changes its orientation, while spacetimes with $0<y<y_{Z,max}(a^2)$ (NS-Z3{\scriptsize C}) contain three such orbits. 
\paragraph{Type D:\quad $1.212<a^2\leq 1.367$}(Fig.~\ref{Fig:3z}D)\\
The function $y_Z(r;a^2)$ has two extrema and $y_h(r;a^2)$ has no extrema; $y_{Z,min}\leq 0$, $y_{Z,max}>0$.
Spacetimes with  $y>y_{Z,max}(a^2)$ (NS-Z1{\scriptsize D}) contain one circular orbit where the centrifugal force vanishes 
independently of the velocity and changes its orientation, while spacetimes with $0<y<y_{Z,max}(a^2)$ (NS-Z3{\scriptsize D})
 contain three such orbits. Spacetimes with $y=y_{Z,max}(a^2)$ (NS-Z2{\scriptsize D}) contain two orbits where the centrifugal 
force vanishes independently of the velocity, but the orientation is changed only on the inner orbit.
\paragraph{Type E:\quad $1.367<a^2<1.813$}(Fig.~\ref{Fig:3z}E)\\
The function $y_Z(r;a^2)$ has two extrema and $y_h(r;a^2)$ has no extrema; $0<y_{Z,min}(a^2)<y_{Z,max}(a^2)$.
Spacetimes with  $y>y_{Z,max}(a^2)$ or $0<y<y_{Z,min}(a^2)$ (NS-Z1{\scriptsize E}) contain one circular orbit where the 

centrifugal force vanishes independently of the velocity and changes its orientation, while spacetimes with 
$y_{Z,min}(a^2)<y<y_{Z,max}(a^2)$ (NS-Z3{\scriptsize E}) contain three such orbits. Spacetimes with $y=y_{Z,min}(a^2)$ 
or $y=y_{Z,max}(a^2)$ (NS-Z2{\scriptsize E}) contain two orbits where the centrifugal force  force vanishes independently 
of the velocity, but the orientation is changed only on one of these orbits.  
\paragraph{Type F:\quad $1.813\leq a^2<\infty$}(Fig.~\ref{Fig:3z}F)\\
The functions $y_Z(r;a^2)$ and $y_h(r;a^2)$ have no extrema.
All the spacetimes with $0<y<\infty$ (NS-Z1{\scriptsize F}) 
contain only one circular orbit where the centrifugal force vanishes independently of the velocity and changes its orientation.\\

\begin{figure}[pb]
\centerline{\psfig{file=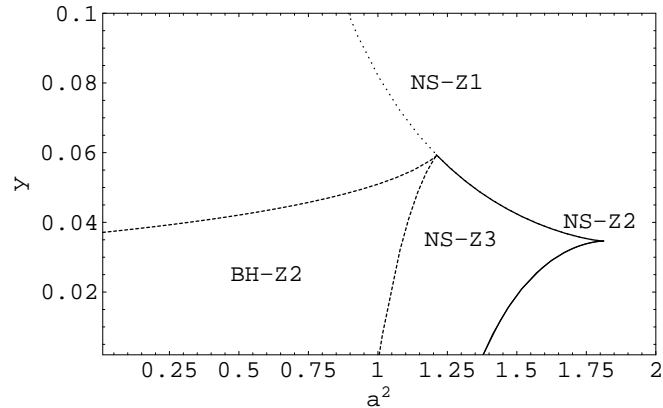,width=0.7\hsize}}
\vspace*{8pt}
\caption{{}The classification of the KdS spacetimes according to the number of circular orbits where 
$\mathcal{Z}(r;a^2,y)=0$. The functions $y_{h,max}(a^2)$ (dashed-upper) and $y_{h,min}(a^2)$ (dashed-lower) 
determining the extrema of $y_h(r;a^2)$ separate regions corresponding to the black-hole spacetimes BH-Z2 with 
two circular orbits where $\mathcal{Z}(r;a^2,y)=0$ and to the naked-singularity spacetimes.  For $a^2>1.212$, the functions 
$y_{Z,max}(a^2)$ (solid-upper) and $y_{Z,min}(a^2)$ (solid-lower) determining the extrema of $y_Z(r;a^2)$ 
determine the spacetimes NS-Z2 with two circular orbits and separate the regions corresponding to the \mbox{NS-Z1} and \mbox{NS-Z3} 
naked-singularity spacetimes with one and three circular orbits where $\mathcal{Z}(r;a^2,y)=0$. For $a^2\leq 1.212$, 
the function $y_{Z,max}(a^2)$ (dotted) does not influence the classification as well as the negative parts of the functions 
$y_{h,min}(a^2)$ and $y_{Z,min}(a^2)$ which are not illustrated here.}
\label{Fig:4z}
\end{figure}

\begin{figure}[pb]
\centerline{\psfig{file=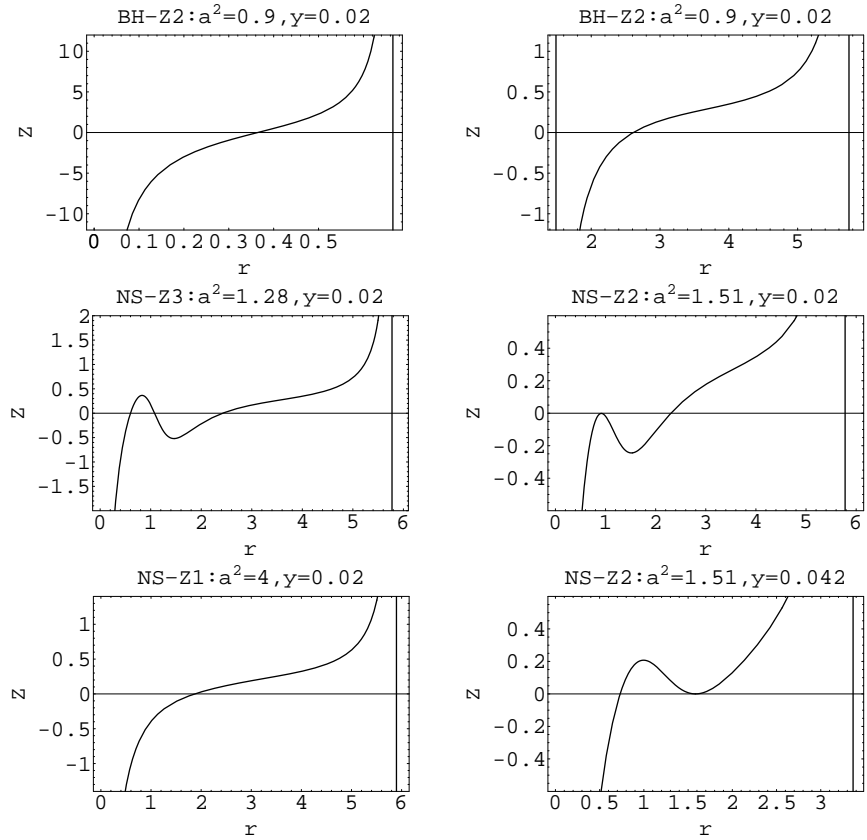,width=\hsize}}
\vspace*{8pt}
\caption{The mass and velocity independent part of the centrifugal force. There are examples of qualitatively different types of behavior of $\mathcal{Z}(r;a^2,y)$ corresponding to the classification of the KdS spacetimes according to the number of circular orbits where  $\mathcal{Z}(r;a^2,y)=0$. The first type of behavior corresponding to the class BH-Z2 is illustrated in the inner and outer stationary black-hole regions separately. Positive (negative) parts of $\mathcal{Z}(r;a^2,y)$ determine repulsive (attractive) force oriented outwards (towards) the singularity at $r=0$. The zero points of the function determine the radii of circular orbits where the centrifugal force vanishes independently of the velocity. The vertical lines denote the asymptotic behavior of the function at event horizons.}
\label{Fig:5z}
\end{figure}

In the parameter plane $(a^2,y)$ of the KdS spacetimes, the functions  $y_{Z,min}(a^2)$, $y_{Z,max}(a^2)$, 
$y_{h,max}(a^2)$, and $y_{h,min}(a^2)$ separate four regions corresponding to one class of the KdS black-hole 
spacetimes BH-Z2\\ 
\begin{center}
\begin{tabular}{cccll}
\hline
\bf{Class}&\bf{Subclass}&\bf{Orbits}&\bf{Interval of} $\mathbf{a^2}$&\bf{Interval of} $\mathbf{y}$\\
\hline
BH-Z2&BH-Z2\scriptsize{A}&2&$(0,1\rangle$&$(0,y_{h,max}(a^2))$\\
&BH-Z2\scriptsize{B}&2&$(1,1.212)$&$\langle y_{h,min}(a^2),y_{h,max}(a^2))$\\
\hline
\end{tabular}
\end{center}
and to three classes of the KdS naked-singularity spacetimes NS-Z1, NS-Z2, and NS-Z3 differing in the number of 
circular orbits where $\mathcal{Z}(r;a^2,y)=0$. 
\begin{center}
\begin{tabular}{cccll}
\hline
\bf{Class}&\bf{Subclass}&\bf{Orbits}&\bf{Interval of} $\mathbf{a^2}$&\bf{Interval of} $\mathbf{y}$\\
\hline
NS-Z1&NS-Z1\scriptsize{A}&1&$(0,1\rangle$&$\langle y_{h,max}(a^2),\infty)$\\ 
&NS-Z1\scriptsize{B}&1&$(1,1.212)$&$\langle y_{h,max}(a^2),\infty)$\\
&NS-Z1\scriptsize{C}&1&$a^2=1.212$&$\langle y_{Z,max}(a^2),\infty)$\\
&NS-Z1\scriptsize{D}&1&$(1.212,1.367\rangle$&$(y_{Z,max}(a^2),\infty)$\\

&NS-Z1\scriptsize{E}&1&$(1.367,1.813)$&$(0,y_{Z,min}(a^2))\cup (y_{Z,max}(a^2),\infty)$\\
&NS-Z1\scriptsize{F}&1&$\langle 1.813,\infty)$&$(0,\infty)$\\  
NS-Z3&NS-Z3\scriptsize{B}&3&$(1,1.212)$&$(0,y_{h,min}(a^2))$\\ 
&NS-Z3\scriptsize{C}&3&$a^2=1.212$&$(0,y_{Z,max}(a^2))$\\ 
&NS-Z3\scriptsize{D}&3&$(1.212,1.367\rangle$&$(0,y_{Z,max}(a^2))$\\ 
&NS-Z3\scriptsize{E}&3&$(1.367,1.813)$&$(y_{Z,min}(a^2),y_{Z,max}(a^2))$\\
NS-Z2&NS-Z2\scriptsize{D}&2&$(1.212,1.367\rangle$&$y=y_{Z,max}(a^2)$\\
&NS-Z2\scriptsize{E}&2&$(1.367,1.813)$&$y=y_{Z,min}(a^2) \cup y_{Z,max}(a^2)$\\  
\hline
\end{tabular}
\end{center} 
In Fig.~\ref{Fig:4z} we give the behavior of the functions  $y_{Z,min}(a^2)$, $y_{Z,max}(a^2)$, $y_{h,max}(a^2)$, and 
$y_{h,min}(a^2)$. Qualitatively different types of behavior of the function $\mathcal{Z}(r;a^2,y)$ corresponding to the given classification are illustrated 
in Fig.~\ref{Fig:5z}. 
\subsubsection*{Schwarzschild-de~Sitter and Kerr cases}
In the SdS spacetimes \cite{Stu:1990:BULAI:} ($a^2=0$, $y>0$), the relation~(\ref{62}) implies the mass and velocity 
independent part of the centrifugal force in the form 
\begin{eqnarray}
\label{82z}
\mathcal{Z}_{SdS}(r;y)=\frac{r-3}{\Delta_r},
\end{eqnarray}
where $\Delta_r=r^2-2r-yr^4$. The centrifugal force is well defined in the stationary regions where \mbox{$\Delta_r>0$}, i.e., at 
all radii between the black-hole and cosmological event horizons given by solutions of the equation~(\ref{83}).   
The function $\mathcal{Z}_{SdS}(r;y)$ diverges at the radii of horizons and vanishes only at the radius $r=3$. Since both horizons coincide at $r=3$ in the case of $y=1/27$, then there is only one class of the stationary SdS spacetimes BH{\scriptsize{SdS}}-Z1 ($0<y<1/27$) with one circular orbit at $r=3$ where the centrifugal force vanishes independently of the velocity and changes its orientation (see Fig.~\ref{Fig:51}).     

In the Kerr spacetimes \cite{Stu-Hle:1999:ACTPS2:} ($a^2>0$, $y=0$), the relation~(\ref{62}) implies the mass and velocity independent part of 
the centrifugal force in the form 
\begin{eqnarray}
\label{85z}
\mathcal{Z}_{K}(r;a^2)=\frac{r^4(r-3)+ra^2[r(r-3)+6]-2a^4}{r\Delta_r[(r+2)a^2+r^3]},
\end{eqnarray}
where $\Delta_r=r^2-2r+a^2$. In the Kerr spacetimes, the centrifugal force is well defined in the stationary region where $\Delta_r>0$, i.e., it is not defined in the dynamic region between the inner and outer black-hole event horizons determined by solutions of the Eq.~(\ref{86}).
It follows from the relation~(\ref{85z}) that for given values of the rotational parameter $a^2$, the centrifugal force 
vanishes independently of the velocity on circular orbits with the radii determined by solutions of the equation
\begin{eqnarray}
\label{87z}
r^4(r-3)+ra^2[r(r-3)+6]-2a^4=0,
\end{eqnarray}
which we consider as the implicit form of the function $a^2_{ZK}(r)$, and restricted by the condition 
\begin{eqnarray}
\label{88z}
a^2_{ZK}(r)>a^2_{hK}(r).
\end{eqnarray}
The function $a^2_{ZK}(r)$ clearly coincides  with the function $a^2_{Z0}(r)$. Its maximum is located at $r\doteq0.812$ and takes 
the value $a^2_{ZK,max}\doteq1.367$. For given values of the rotational parameter $a^2$, the local extrema of the function 
$\mathcal{Z}_{K}(r;a^2)$ are given by solutions of the equation 
\begin{eqnarray}
\label{89z}
\nonumber 4a^8(1+r)+a^6r\{r[r(20+r)-12]-16\}+a^4r^2\{24+r^2[r(26+r)-78]\}-\\
a^2r^5[r(r-6)(r-10)-72]-r^8[r(r-6)+6]=0,
\end{eqnarray}
which we consider as the implicit form of the function $a^2_{ZeK}(r)$. The radii of circular orbits where the centrifugal force vanishes independently of the velocity and does not change its orientation are determined by common points of $a^2_{ZK}(r)$ and $a^2_{ZeK}(r)$. The only  such point satisfying the condition~(\ref{88z}) corresponds to the maximum 
of $a^2_{ZK}(r)$.  
 
\begin{figure}[pb]
\centerline{\psfig{file=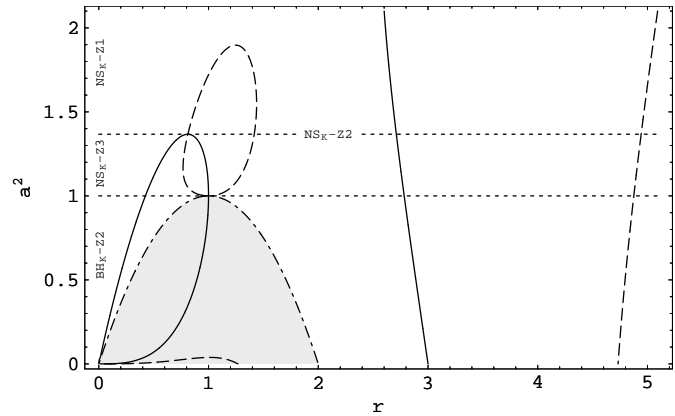,width=0.7\hsize}}
\vspace*{8pt}
\caption{Characteristic functions $a^2_{hK}(r)$ (dashed-dotted) determining the loci of event horizons and limiting the 
dynamic region where $a^2<a^2_{hK}(r)$ (gray); $a^2_{ZK}(r)$ (solid) determining the circular orbits where 
$\mathcal{Z}_{K}(r;a^2)=0$; $a^2_{ZeK}(r)$ (dashed) determining the extrema of $\mathcal{Z}_{K}(r;a^2)$.  
For given $a^2$, the loci of horizons are determined by the solutions of $a^2=a^2_{hK}(r)$. The circular orbits where 
$\mathcal{Z}_{K}(r;a^2)=0$ and the extrema of $\mathcal{Z}_{K}(r;a^2)$, respectively, are determined by the solutions of 
$a^2=a^2_{ZK}(r)$ and $a^2=a^2_{ZeK}(r)$, respectively, and restricted by the conditions $a^2_{ZK}(r)>a^2_{hK}(r)$ and $a^2_{ZeK}(r)>a^2_{hK}(r)$ (i.e., the parts of $a^2_{ZK}(r)$ and $a^2_{ZeK}(r)$ in the gray region or on its border are irrelevant).  
In the parameter line ($a^2$), the extrema of the characteristic functions separate regions corresponding to one 
class of the Kerr black-hole spacetimes BH{\scriptsize{K}}-Z2 with two circular orbits and to three classes of the 
Kerr naked-singularity spacetimes NS{\scriptsize{K}}-Z3, NS{\scriptsize{K}}-Z2, and NS{\scriptsize{K}}-Z1 with 
(subsequently) three, two, and one circular orbits where $\mathcal{Z}_{K}(r;a^2)=0$.}
\label{Fig:6z}
\end{figure}

\begin{figure}[pb]
\centerline{\psfig{file=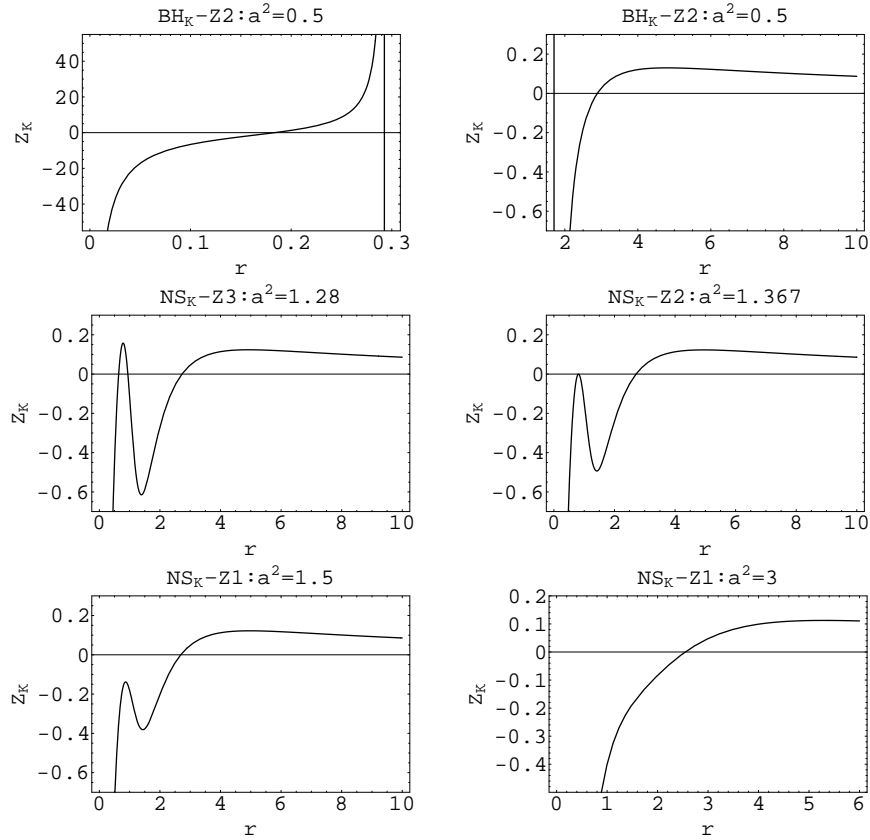,width=\hsize}}
\vspace*{8pt}
\caption{The mass and velocity independent part of the centrifugal force. There are examples of qualitatively different types of behavior of $\mathcal{Z}_K(r;a^2)$ corresponding to the classification of the Kerr spacetimes according to the number of circular orbits where $\mathcal{Z}_K(r;a^2)=0$. The first type of behavior corresponding to the class BH{\scriptsize{K}}-Z2  is illustrated in the inner and outer stationary black-hole regions separately. Positive (negative) parts of $\mathcal{Z}_K(r;a^2)$ determine repulsive (attractive) force oriented outwards (towards) the singularity at $r=0$. The zero points of the function determine the radii of circular orbits where the centrifugal force vanishes independently of the velocity. The vertical lines denote the asymptotic behavior of the function at event horizons.}
\label{Fig:7z}
\end{figure}

All the characteristic functions $a^2_{hK}(r)$, $a^2_{ZK}(r)$, and $a^2_{ZeK}(r)$ are illustrated in Fig.~\ref{Fig:6z}.
In the parameter line ($a^2$) of the Kerr spacetimes, the extrema of the characteristic functions separate regions corresponding to one class of the Kerr black-hole spacetimes BH{\scriptsize{K}}-Z2 and to three classes of naked-singularity spacetimes NS{\scriptsize{K}}-Z1, NS{\scriptsize{K}}-Z2, and NS{\scriptsize{K}}-Z3 differing in the number of circular orbits where $\mathcal{Z}_{K}(r;a^2)=0$.\quad\\
\paragraph{Class BH{\scriptsize{K}}-Z2:\quad $0<a^2\leq 1$}\quad\\
In the inner stationary region of the spacetime, there is one orbit where the centrifugal force vanishes independently 
of the velocity and changes its orientation as well as in the outer stationary region.
\paragraph{Class NS{\scriptsize{K}}-Z3:\quad $1<a^2<1.367$}\quad\\
There are three orbits where the centrifugal force vanishes independently of the velocity and changes its orientation. 
\paragraph{Class NS{\scriptsize{K}}-Z2:\quad $a^2=1.367$}\quad\\
There are two orbits where the centrifugal force vanishes independently of the velocity, but the orientation is changed 
only on the outer orbit.
\paragraph{Class NS{\scriptsize{K}}-Z1:\quad $a^2>1.367$}\quad\\
There is one orbit where the centrifugal force vanishes independently of the velocity and changes its orientation. 
\\\\
Qualitatively different types of behavior of the function $\mathcal{Z}_{K}(r;a^2)$    corresponding to the given classification are illustrated in \mbox{Fig.~\ref{Fig:7z}.}

\section{Conclusions}

The Kerr-de~Sitter black-hole spacetimes contain two stationary regions in the equatorial plane. The inner region is limited 
by singularity and by the inner black-hole horizon. The outer region is limited by the outer horizon and by the cosmological 
horizon. For any given values of the rotational parameter $a$ and the cosmological parameter $y$, in each of these regions, 
there is only one circular orbit where the gravitational force vanishes, and only one orbit where the centrifugal force 
vanishes independently of the velocity. Both the gravitational and centrifugal forces change their orientations on these 
orbits. The same situation occurs in the Kerr black-hole spacetimes \cite{Stu-Hle:1999:ACTPS2:}, except for the outer 
stationary region, where there is no circular orbit where the gravitational force vanishes, in contrast to the 
Kerr-de~Sitter outer stationary region. In the only stationary region of the Schwarzschild-de~Sitter spacetimes 
\cite{Stu:1990:BULAI:}, there is also only one  circular orbit where the gravitational force vanishes and one circular 
orbit where the centrifugal force vanishes independently of the velocity. 

The Kerr-de~Sitter naked-singularity spacetimes contain one stationary region between the singularity and the cosmological 
horizon. In some of these spacetimes, even three circular orbits where the gravitational force vanishes and other three orbits 
where the centrifugal force vanishes independently of the velocity can occur. This indicates a relatively complex structure 
of these spacetimes as a result of mixed influence of the rotation of the source and the cosmological repulsion. It is more 
complicated situation than in the Kerr naked-singularity spacetimes, where the maximum number of orbits where the gravitational force vanishes is only two. On the other hand, there are at most three circular orbits where the centrifugal force vanishes independently of the velocity in accord with the case of the Kerr-de~Sitter naked-singularity spacetimes \cite{Stu-Hle:1999:ACTPS2:}.

Clearly, in the Kerr-de~Sitter black-hole spacetimes as well as in the naked-singularity spacetimes, the additional radii where the gravitational force vanishes (compared to the Kerr spacetimes) occur due to the effect of the cosmological repulsion.

\section*{Acknowledgments}

This work was supported by the Czech grant MSM 4781305903 and by the Committee for collaboration of the Czech Republic with CERN. Z.Stuchl\'{i}k would like to thank Theory Division of CERN for perfect hospitality.


\begin{thebibliography}{00}

\bibitem{Abr-Car-Las:1988:GENRG2:}
M.~A. Abramowicz, B.~Carter and {J.P.} Lasota, 
{\it Gen. Relativity Gravitation} {\bf 20}, 1173 (1988).

\bibitem{Abr:1990:MONNR:}
M.~A. Abramowicz, 
{\it Monthly Notices Roy. Astronom. Soc.} {\bf 245(4)}, 733 (1990).

\bibitem{Abr-Nur-Wex:1993:CLAQG:}
M.~A. Abramowicz, P.~Nurowski and N.~Wex,
{\it Classical Quantum Gravity} {\bf 10}, L183 (1993). 

\bibitem{Abr-Nur-Wex:1995:CLAQG:}
M.~A. Abramowicz, P.~Nurowski and N.~Wex, 
{\it Classical Quantum Gravity} {\bf 12(6)}, 1467 (1995).

\bibitem{Jan-Car-Bin:1992:ANNPH1:}
R.~T. Jantzen, P.~Carini and D.~Bini, 
{\it Ann. Physics} {\bf 215(1)}, 1 (1992). 

\bibitem{Abr-Mil:1990:MONNR:}
M.~A. Abramowicz and J.~C. Miller, 
{\it Monthly Notices Roy. Astronom. Soc.} {\bf 245(4)}, 729 (1990).

\bibitem{Stu:1990:BULAI:}
Z.~Stuchl\'{i}k,
{\it Bull. Astronom. Inst. Czechoslovakia} {\bf 41}, 341 (1990).

\bibitem{Abr:1992:MONNR:}
M.~A. Abramowicz,
{\it Monthly Notices Roy. Astronom. Soc.} {\bf 256(4)}, 710 (1992).

\bibitem{Abr-Mil-Stu:1993:PHYSR4:}
M.~A. Abramowicz, J.~Miller and Z.~Stuchl\'{i}k,
{\it Phys. Rev. D} {\bf 47(4)}, 1440 (1993).

\bibitem{Abr:1993:RenGRCos:}
M.~A. Abramowicz,
{\it The Renaissance of General Relativity and Cosmology} (Cambridge, 1993), p.~ 40.

\bibitem{Abr-etal:1997:CLAQG:}
M.~A. Abramowicz, N.~Anderson, M.~Bruni, P.~Ghosh and S.~Sonego,
{\it Classical Quantum Gravity} {\bf 14}, L189 (1997).

\bibitem{Abr-etal:1997:GENRG2:}
M.~A. Abramowicz, A.~Lanza, J.~C. Miller and S.~Sonego,
{\it Gen. Relativity Gravitation} {\bf 29(12)}, 1585 (1977).

\bibitem{Abr:1999:PHYSR:}
M.~A. Abramowicz,
{\it Phys. Rep.} {\bf 311}, 325 (1999).

\bibitem{Kri-Son-Abr:1998:GENRG2:}
S.~Kristiansson, S.~Sonego and M.~A. Abramowicz,
{\it Gen. Relativity Gravitation} {\bf 30(2)}, 275 (1998).

\bibitem{Stu-Hle:1999:ACTPS2:}
Z.~Stuchl\'{i}k and S.~Hled\'{i}k,
{\it Acta Phys. Slovaca} {\bf 49(5)}, 795 (1999).

\bibitem{Stu-Hle:1999:CLAQG:}
Z.~Stuchl\'{i}k and S.~Hled\'{i}k,
{\it Classical Quantum Gravity} {\bf 16(4)}, 1377 (1999).

\bibitem{Stu-Hle-Jur:2000:CLAQG:}
Z.~Stuchl\'{i}k, S.~Hled\'{i}k and J.~Jur\'{a}\v{n},
{\it Classical Quantum Gravity} {\bf 17(14)}, 2691 (2000).

\bibitem{Hle:2001:RAGtime2and3:}
S.~Hled\'{i}k,
{\it Proc. of RAGtime
  2/3: Workshops on black holes and neutron stars 11--13/8--10 October
  2000/01}, (Czech Republic, Opava, 2001), p.~25-52. 

\bibitem{Stu-Hle:1999:PHYSR4:}
Z.~Stuchl\'{i}k and S.~Hled\'{i}k,
{\it Phys. Rev. D} {\bf 60(4)}, 044006 (1999).

\bibitem{Stu-Hle:2002:ACTPS2:}
Z.~Stuchl\'{i}k, S.~Hled\'{i}k,
{\it Acta Phys. Slovaca} {\bf 52(5)}, 363 (2002).

\bibitem{Stu-Sla:2004:PHYSR4:}
Z.~Stuchl\'{i}k and P.~Slan\'{y},
{\it Phys. Rev. D} {\bf 69}, 064001 (2004).

\bibitem{deFel:1974:ASTRA:}
F.~de~Felice,
{\it Astronomy and Astrophysics} {\bf 34}, 15 (1974).

\end{thebibliography}
\end{document}